\documentclass[pdflatex,sn-mathphys-num,iicol]{sn-jnl}

 \usepackage{xr} 
\usepackage{graphicx}%
\usepackage{amsmath,amssymb,amsfonts}%

\title{Subdiffusion from competition between multi-exponential friction memory and energy barriers}

\author[1]{\fnm{Anton} \sur{Klimek}}

\author[1]{\fnm{Benjamin A.} \sur{Dalton}}

\author*[1]{\fnm{Roland R.} \sur{Netz}}\email{rnetz@physik.fu-berlin.de}

\affil[1]{\orgdiv{Fachbereich Physik}, \orgname{Freie Universit\"at Berlin}, \orgaddress{\street{Arnimalle 14}, \city{Berlin}, \postcode{14195}, \country{Germany}}}

\begin{document}

    \abstract{Subdiffusion is a hallmark of complex systems, ranging from protein folding to transport in viscoelastic media.
    However, despite its pervasiveness, the mechanistic origins of subdiffusion remain contested.
    Here, we analyze both Markovian and non-Markovian dynamics, in the presence and absence of energy barriers, in order to disentangle the distinct contributions of memory-dependent friction and energy barriers to the emergence of subdiffusive behavior.
    Focusing on the mean squared displacement (MSD), we develop an analytical framework that connects subdiffusion to multiscale memory effects in the generalized Langevin equation (GLE), and derive the subdiffusive scaling behavior of the MSD for systems governed by multi-exponential memory kernels.
    We identify persistence and relaxation timescales that delineate dynamical regimes in which subdiffusion arises from either memory or energy barrier effects.
    By comparing analytical predictions with simulations, we confirm that memory dominates the overdamped dynamics for barrier heights up to approximately $2\,k_BT$, a regime recently shown to be relevant for protein folding.
    Overall, our results advance the theoretical understanding of anomalous diffusion and provide practical tools that are broadly applicable to fields as diverse as molecular biophysics, polymer physics, and active matter systems.}

    \maketitle
	
	\section{Introduction}
    Diffusion is a fundamental transport phenomenon observed across a wide range of physical, chemical, and biological systems.
    It is driven by interactions of an observable with its fluctuating environment, which results in random motion.
    Therefore, the dynamics of a single diffusing observable are described by a random walk and its statistics are captured by ensemble- or time averages, which are equivalent for stationary systems \cite{jack2020ergodicity}.
    An important statistical description of the dynamics of a system is the MSD, defined as $C_{\textrm{MSD}}(t) = \langle (x(0) - x(t))^2 \rangle$, where $\langle \cdot \rangle$ denotes an ensemble average and $x$ is the observable of interest.
    $C_{\textrm{MSD}}(t)$ characterizes the diffusion dynamics of an observable and can reveal important information about the physical properties of the environment.
    In general, the MSD can be written in the form $C_{\rm{MSD}}(t)\propto t^{\alpha(t)}$ with the time dependent exponent $\alpha(t)$ describing the mode of diffusion at a given time.
    The most common model of diffusion is characterized by a linear increase of the MSD in time, $\alpha(t)=1$, and is referred to as Brownian- or normal diffusion.
    Brownian diffusion can describe the dynamics of a molecule in a liquid \cite{einstein_neue_1906}, but also the dynamics of stock prices \cite{kiefer_predictability_2024} or the motion of certain cells \cite{gail_locomotion_1970}.
    However, the dynamics of many physical, chemical, and biological systems are known to deviate from normal diffusive behavior \cite{bouchaud1990anomalous,klages2008anomalous,lindner2010diffusion,sokolov2012models}.
    The mode of diffusion of an observable is characterized by the time-dependent exponent, given by
	\begin{equation}
		\label{eq_alpha}
		\alpha(t) = \frac{{d} \ln\big(C_{\text{MSD}}(t)\big)}{{d} \ln(t)}\,,
	\end{equation}
    where $\alpha(t)<1$ is called subdiffusion, $\alpha(t)>1$ superdiffusion, the special case $\alpha(t)=1$ defines normal diffusion and $\alpha(t)=2$ defines ballistic motion.
    It is often the case that for a given observable, $\alpha(t)$ is not constant but rather varies significantly as a function of $t$.
    As such, systems often exhibit different modes of diffusion on different time scales.
    
	Examples of superdiffusion include the motion of animals \cite{viswanathan_physics_2011}, active particles \cite{douglass_superdiffusion_2012} and actively transported particles within cells \cite{reverey_superdiffusion_2015} and are often related to non-equilibrium phenomena.
    Subdiffusion, on the other hand, appears in crowded, caged or viscoelastic environments and for systems with energy barriers in the potential landscape.
	Examples of subdiffusion include cell migration \cite{dieterich_anomalous_2008,dieterich_anomalous_2022}, polymer network dynamics \cite{guo_entanglement-controlled_2012,abbasi_non-markovian_2023}, biomolecule diffusion in crowded environments \cite{berry_anomalous_2014,weiss_anomalous_2004,hansing2016nanoparticle}, protein folding dynamics \cite{sangha_proteins_2009,chung_structural_2015,baldwin_is_1999} and many more.
    The interplay between energy barriers and viscoelastic effects, that both contribute to deviations from normal diffusion, is still not well understood.

	Theoretical approaches to describe subdiffusive phenomena use diverse models such as correlated time random walks \cite{magdziarz_fractional_2009,jeon_noisy_2013,cairoli_langevin_2015} or fractal Brownian motion \cite{gmachowski_fractal_2015,ernst_fractional_2012}.
	Many of these models are special cases of the generalized Langevin equation (GLE), which is derived exactly from a many-body Hamiltonian via projection \cite{mori_transport_1965,ayaz_generalized_2022,mitterwallner_negative_2020}.
	The GLE incorporates non-Markovian effects, i.e. a trajectory's memory of its past, which naturally arises when projecting the dynamics of a many-body system on a low- or one-dimensional observable of interest \cite{mori_transport_1965,ayaz_generalized_2022,dalton2024memory}.
	
	The GLE describes the motion of a general observable $x$ in a complex environment as
    \begin{equation}
        \label{eq_gle_wom}
        \ddot{x}(t)=-\nabla U(x(t)) -\int_{0}^{t} \Gamma(t-t')\dot{x}(t') dt' +F_R(t) \,,
    \end{equation}
    where $\Gamma(t)$ is the friction memory kernel that describes the dependence of the friction force on previous velocities, $-\nabla U(x)$ is the force due to a potential, and $F_R(t)$ is a random force that incorporates the fluctuations of the environment.
    Note that no mass appears in front of the acceleration term, as is appropriate for a general description including equilibrium and non-equilibrium systems, so that all terms in Eq.~\eqref{eq_gle_wom} have units of acceleration.
    The correlation of the random forces using the Mori projection in a harmonic potential $U(x)$ \cite{mori_transport_1965} fulfills the relation
	\begin{equation}
		\label{eq_fdt}
		\langle F_R(0) F_R(t) \rangle = B \Gamma (t)\,,
	\end{equation}
	where $B=\langle \dot{x}^2\rangle$, which holds only approximately for non-harmonic potentials $U(x)$ \cite{vroylandt_derivation_2022, ayaz_generalized_2022}.
	A GLE Eq.~\eqref{eq_gle_wom} can also be derived for non-equilibrium systems \cite{netz_derivation_2024} with an expression for the random force fluctuations that is more general than Eq.~\eqref{eq_fdt}.
    Such non-equilibrium GLEs can be used to describe cell motion \cite{mitterwallner_non-markovian_2020,klimek_data-driven_2024,klimek_intrinsic_2024} or other actively driven processes \cite{abbasi_non-markovian_2023}.
	
	Models that describe subdiffusive phenomena within the GLE framework often assume power-law memory $\Gamma(t)\propto t^{-\alpha}$ motivated by fractal patterns and self similarity of the systems under consideration \cite{kou_generalized_2004,metzler_random_2000,kappler_cyclization_2019}.
	It was shown that power-law memory can efficiently be modeled by sums of exponential memory components \cite{noauthor_advances_2012,goychuk_viscoelastic_2009}.
	In fact, protein folding dynamics observed in molecular dynamics (MD) simulations have been described using multi-exponential memory, which accurately captures the subdiffusive behavior in protein dynamics \cite{ayaz_non-markovian_2021,dalton_fast_2023}.

    In this paper, we investigate how multi-exponential memory gives rise to subdiffusivity, characterized by a scaling exponent ${\alpha < 1}$ in the MSD.
    We introduce an analytical framework for predicting the MSD for non-Markovian systems with multi-exponential memory kernels, which is exactly solvable for free diffusion and diffusion in a harmonic potential.
    We compare the analytical results with non-Markovian simulations in a double-well potential to determine the respective roles of memory-dependent friction and energy barriers in determining subdiffusive dynamics.
    Motivated by the exponential memory observed in protein folding \cite{ayaz_non-markovian_2021, dalton_fast_2023, dalton2024ph}, we use multi-exponential memory with exponentially spaced memory times and amplitudes and derive an exact expression for the scaling exponent $\alpha$ of the MSD.
    We find that memory dictates the dynamics on short and intermediate time scales, and we give expressions for the time scales on which memory effects dominate over energy barriers in determining the MSD.
    We determine the transition time between ballistic and subdiffusive behavior and show that the ballistic regime of the MSD is extended under the influence of memory.
    Furthermore, we show that for small energy barriers (on the order of a few $k_\text{B}T$), the influence of the potential landscape on the MSD is negligible in comparison to the influence of memory effects, until achieving time scales at which the system probes the global spatial confinement.
    Overall, our results shed light on the connection between multi-scale memory and subdiffusion, and we demonstrate that under many conditions, memory effects are more important than the potential landscape in generating subdiffusive dynamics.
    
\section{Non-Markovian diffusion model}
	\begin{figure}
		\hspace{-0.5cm}
		\includegraphics{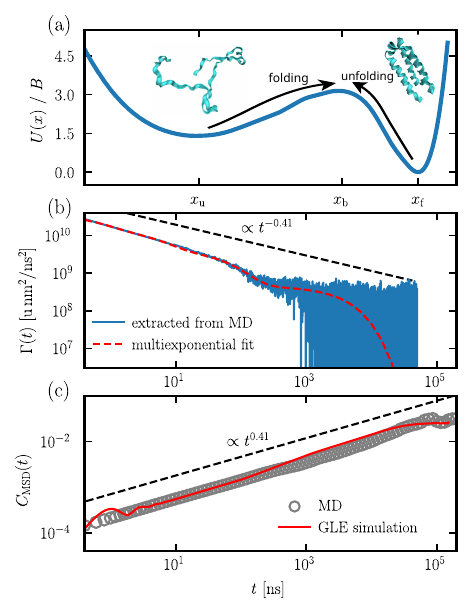}
		\caption{(a) Potential $U(x)$ extracted from MD simulations of the fast-folding protein $\alpha_3$D, originally published in \cite{lindorff-larsen_how_2011}, as a function of the fraction of native contacts reaction coordinate $x$. 
        $x_{\rm{f}}$ and $x_{\rm{u}}$ are the locations of the folded and unfolded states, respectively, and $x_{\rm{b}}$ is the location of the barrier top.
        Representative structures for the folded and unfolded states are shown at their respective positions.
        (b) Friction kernel $\Gamma(t)$ of the fraction of native contacts reaction coordinate for $\alpha_3$D, extracted from MD trajectories, and the corresponding multi-exponential fit of Eq.~\eqref{eq_kern_expo} to the extracted kernel.
        (c) MSD of the fraction of native contacts reaction coordinate of $\alpha_3$D from MD (gray circles).
        The red line shows results for the simulation of the GLE (Eq.~\eqref{eq_gle_wom}) with the fitted friction kernel $\Gamma(t)$ in (b) and the potential $U(x)$ in (a).
        The dashed lines in (b) and (c) display pure power-law behavior with $\alpha$ determined by a logarithmic time average of Eq.\eqref{eq_alpha}, i.e. an average using exponentially spaced time points, over the MD data shown in (c).}
		\label{fig1_motivation}
	\end{figure}
    The friction memory kernel $\Gamma(t)$, which represents dissipative interactions with a dynamic environment, can take different forms depending on the specific details of a particular system.
    In the context of subdiffusive phenomena, two types of kernels are often considered: power-law kernels \cite{kou_stochastic_2008, panja_generalized_2010} and sums of exponentials \cite{ayaz_non-markovian_2021, goychuk_viscoelastic_2009, sprenger_active_2022}.
    Power-law kernels result in subdiffusive behavior across all time scales. 
    However, relaxation times of physical systems are typically bound by both a smallest and a largest time scale.
    It has been shown that the dynamics for systems with multi-exponential memory kernels can give rise to power-law behavior in the corresponding MSDs over many orders of magnitude, with short-time and long-time cutoffs \cite{noauthor_advances_2012, goychuk_viscoelastic_2009}.
    In fact, the dynamics of many fast-folding proteins are well described by a GLE with multi-exponential memory kernels, moving over a potential landscape $U(x)$ \cite{ayaz_non-markovian_2021, dalton_fast_2023, dalton2024ph, dalton_role_2024}.
    To illustrate this, in Fig.~\ref{fig1_motivation}a, we show the potential of mean force extracted from all-atom simulations of the fast-folding protein $\alpha_3$D, and in Fig.~\ref{fig1_motivation}b we show that the multi-exponential memory kernel $\Gamma(t)$ extracted from molecular dynamics simulations of $\alpha_3$D \cite{lindorff-larsen_how_2011, dalton_fast_2023} closely resembles a power-law kernel until an upper cutoff time.
    A GLE simulation can be parameterized using a fit to the memory kernel shown in Fig.~\ref{fig1_motivation}b and the extracted potential $U(x)$ shown in Fig.~\ref{fig1_motivation}a.
    The MSD generated by the GLE simulation can then be compared to the MSD directly extracted from the all-atom MD simulations.
    In Fig.~\ref{fig1_motivation}c, we show that the GLE with multi-exponential memory accurately reproduces the MD simulation MSD.
    Therefore, in this paper, we consider multi-exponential kernels of the form
	\begin{equation}
		\label{eq_kern_expo}
		\Gamma(t) = \sum_{i=1}^{n}\frac{\gamma_i}{\tau_i} e^{-t/\tau_i} \,.
	\end{equation}
    The potential $U(x)$ in Fig.~\ref{fig1_motivation}a exhibits two distinct local minima corresponding to the folded and unfolded states, separated by a barrier of height $U_0=3B$ from the folded state (corresponding to $3\,k_{\text{B}}T$), where the observable $x$ is the fraction of native contacts reaction coordinate \cite{best2013native, dalton_fast_2023}, commonly used for describing the conformational states in protein simulations.
    Energy barriers, such as those observed in Fig.~\ref{fig1_motivation}a, occur in many two- or multi-state systems.
    How such energy barriers influence the MSDs that are measured in dynamic systems and what role the barriers play in comparison to friction memory effects, is still an open question. 
    Recent investigations into protein-folding dynamics \cite{dalton_fast_2023, ayaz_non-markovian_2021} and into the role of power-law kernels fitted by exponentials over finite time intervals \cite{noauthor_advances_2012, goychuk_viscoelastic_2009, goychuk_finite-range_2020} demonstrate the importance of exponential memory contributions in complex systems.
    Motivated by the applicability to protein folding and other complex systems, and aiming to systematically analyze the dynamics of observables that exhibit multi-exponential friction memory kernels, we consider exponentially spaced memory time scales and friction amplitudes, according to
    \begin{align}
            \label{eq_cd}
        \begin{split}
        \tau_i &= \tau_1 c^{i-1} \\
        \gamma_i &= \gamma_1 d^{i-1}\,,
        \end{split}
    \end{align}    
    where $c$ determines the memory-time ratio $\tau_{i+1}/\tau_i$ and $d$ determines the friction-amplitude ratio $\gamma_{i+1}/\gamma_i$.

    In SI Sec. \ref{sec_derivation_alpha}, we derive the scaling exponent $\alpha$ of the MSD for a multi-exponential memory kernel defined by Eqs.~\eqref{eq_kern_expo},~\eqref{eq_cd} as
    \begin{equation}
        \label{eq_alpha_cd_main}
        \alpha(c,d) = \ln(c/d) / \ln(c),
    \end{equation}
    valid for $c \gg d >1$ and for times $\tau_1<t<\tau_n$.
    We verify Eq.~\eqref{eq_alpha_cd_main} further below by comparison with our analytic and simulation results.
	
    It is useful to define the running integral of the friction kernel as
    \begin{equation}
        \label{eq_int_kern}
        G(t) = \int_{0}^{t} \Gamma(s) ds\,.
    \end{equation}
    From this, the inertial time is defined as
    \begin{equation}
        \label{eq_total_friction}
        \tau_m = \frac{1}{G(\infty)} = \frac{1}{\sum_{i=1}^n \gamma_i}\,,
    \end{equation}
    which governs the long-time diffusivity via $D=B\tau_m$.
    In the Markovian limit, where all memory times are small compared to the inertial time ($\tau_i \ll \tau_m$), one recovers the Markovian Langevin equation \cite{kappler_memory-induced_2018, wisniewski_memory_2024} given by
    \begin{equation}
        \label{eq_le_wom}
        \ddot{x}(t) = -\nabla U(x(t)) - \frac{1}{\tau_m}\dot{x}(t)  + F_R(t)\,,
    \end{equation}
    which leads to dynamics with a short-time ballistic regime in the MSD with $\alpha=2$, directly followed by long-time diffusive motion with $\alpha=1$.
    In the case of free diffusion ($U(x)=0$), the diffusive regime continues unbounded, whereas potentials of finite width lead to confinement, with a long-time plateau in the MSD and correspondingly $\alpha=0$ for long times.
    The timescale at which the ballistic regime terminates defines the persistence time $\tau_p$, which is $\tau_p=\tau_m$ in the Markovian case with $\tau_i \ll \tau_m$.
    As we discuss below, in the presence of longer memory times $\tau_i > \tau_m$, the persistence time $\tau_p$ is prolonged such that $\tau_p>\tau_m$.
    
    \section{Results \& Discussion}
    We study the effects of multi-exponential memory on the MSD in the absence of potentials, in the presence of harmonic confinement without energy barriers, and in double-well potentials with energy barriers.
    The comparison of the MSD in the presence and absence of memory effects in different potential landscapes allows us to determine the individual influence of memory, confinement, and energy barriers on the dynamics.

	\subsection{Free diffusion}
    In SI Sec.~\ref{sec_msd_expo_derivation}, we derive the analytic result for the MSD for a general multi-exponential kernel Eq.~\eqref{eq_kern_expo} in the presence of a harmonic potential $U(x)=Kx^2/2$ as    
    \begin{align}
        \label{eq_msd}
        &C_{\rm{MSD}}(t) =  \\
        &\frac{B}{c_{n+2}} \left( \sum_{i=1}^{n+2}\frac{e^{-\sqrt{-\omega^2_i} t} - 1}{\sqrt{-\omega^2_i} \prod_{j\neq i} (\omega^2_i-\omega^2_j) }  \sum_{m=1}^{2n-1}k_m \omega_i^{2m-2}  \right)\,, \nonumber
    \end{align}
    where $c_i$, $k_i$, and $\omega_i$ are constants determined by the kernel parameters $\gamma_i$ and $ \tau_i$, and by the harmonic coupling strength $K$. 
    In the SI Sec.~\ref{sec_msd_expo_derivation}, we give explicit examples of Eq.~\eqref{eq_msd} for $n=3$ and $n=5$. 
    To understand the impact of friction memory $\Gamma(t)$ on the MSD, it is instructive to first analyze GLE dynamics in the absence of an external potential, i.e., $U(x)=0$.
    
\begin{figure*}
		\includegraphics{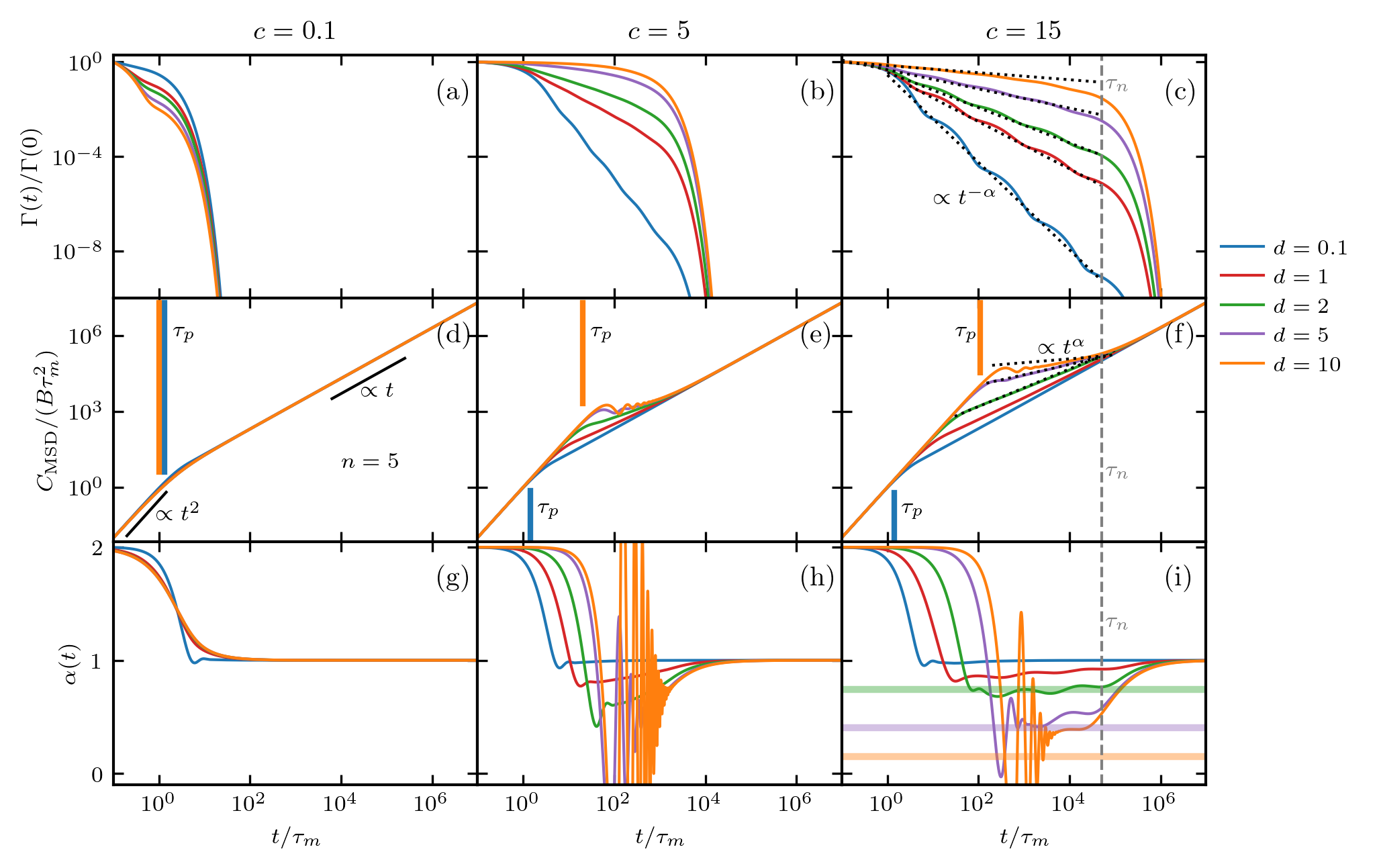}
		\caption{Non-Markovian dynamics in the absence of a potential ($U(x)=0$).
		Results are shown for $n=5$ and $\tau_1=\tau_m$.
		(a)-(c) Multi-exponential memory kernel $\Gamma(t)$ (Eq.~\eqref{eq_kern_expo}), for a range of $c$ and $d$ values, as defined by Eq.~\eqref{eq_cd}.
		The dotted lines in (c) represent the power law predicted by Eq.~\eqref{eq_alpha_cd_main}.
		(d)-(f) Analytic results for the MSD given by Eq.~\eqref{eq_msd}, determined by the friction kernels given in (a)-(c).
		The colored vertical lines show the persistence time $\tau_p$ for $d=0.1$ and $d=10$, defined in Eq.~\eqref{eq_def_taup}.
		The black lines in (d) indicate ballistic ($\alpha=2$) and diffusive ($\alpha=1$) scaling behavior.
        The dotted black lines in (f) show the predicted subdiffusive scaling of the MSDs for the kernels in (c) for $c>d>1$, with the predicted $\alpha$ given by Eq.~\eqref{eq_alpha_cd_main}.
		(g)-(i) Time-dependent exponent $\alpha(t)$ of the MSDs in (d)-(f), obtained via Eq.~\eqref{eq_alpha}.
		The horizontal lines in (i) represent the prediction of $\alpha$ by Eq.~\eqref{eq_alpha_cd_main}, where the color scheme matches the curves for $\alpha(t)$.
		The vertical dashed lines in (c), (f), and (i) show the longest memory time $\tau_n$.}
		\label{fig_msd_free_alpha_n5}
	\end{figure*}

    In Fig.\ref{fig_msd_free_alpha_n5}, we present analytical results for MSDs Eq.~\eqref{eq_msd} for $U(x) = 0$, which means $K=0$.
    The input memory kernels $\Gamma(t)$ are shown in Figs.~\ref{fig_msd_free_alpha_n5}a-c, the corresponding MSDs in Figs.~\ref{fig_msd_free_alpha_n5}d-f, and the time-dependent exponents $\alpha(t)$ in Figs.~\ref{fig_msd_free_alpha_n5}g-i.
    In Figs.~\ref{fig_msd_free_alpha_n5}a-c, we demonstrate that for certain combinations of $c$ and $d$, the multi-exponential friction kernels exhibit extended regions following a power law $\Gamma(t)\propto t^{-\alpha}$, where $\alpha$ is predicted by Eq.~\eqref{eq_alpha_cd_main} for $c > d$.
	This is evident in Fig.~\ref{fig_msd_free_alpha_n5}c, where we plot the corresponding power laws using exponents from Eq.~\eqref{eq_alpha_cd_main} as dotted lines.
	In Fig.~\ref{fig_msd_free_alpha_n5}f, we see that the region of the MSD between the short-time ballistic regime and the long-time diffusive regime is also described well by a power law $C_{\rm{MSD}}\propto t^\alpha$ \cite{burov_critical_2008, noauthor_advances_2012}, with the same $\alpha$ that describes the memory kernels, when $\alpha<1$.
	The predicted exponents (Eq.~\eqref{eq_alpha_cd_main}) agree well with $\alpha(t)$ (Eq.~\ref{eq_alpha}) in those regions, as validated by the horizontal lines in Fig.~\ref{fig_msd_free_alpha_n5}i.
	Beyond the longest memory time $\tau_n=\tau_1 c^{n-1}$, the kernels deviate from power-law behavior, clearly displaying an exponential decay for $t>\tau_n$.
    This corresponds to the crossover to the long-time diffusive regime with $\alpha=1$, as can be seen in Figs.~\ref{fig_msd_free_alpha_n5}c,f,i.
    
	In Figs.~\ref{fig_msd_free_alpha_n5}d and g, where $c=0.1$, we observe the direct transition from the ballistic regime ($\alpha=2$) to long-time free diffusion ($\alpha=1$) at the inertial time $t=\tau_m$, as expected for $c<1$ and $d>1$ ($\tau_i<\tau_m$), which represents the Markovian limit.
    As demonstrated in Fig.~\ref{fig_msd_free_alpha_n5}, the ballistic regime extends with increasing $c$ and $d$.
	The persistence time is given by the solution to the self-consistent equation
	\begin{equation}
		\label{eq_def_taup}
		\tau_p = 1 / G(\tau_p)\,,
	\end{equation}
	shown as vertical lines in Figs.~\ref{fig_msd_free_alpha_n5}d-f and determines the end point of the ballistic regime in the MSD.
	The persistence time is the generalization of the inertial time in the case of time-dependent friction $\Gamma(t)$.
	In the absence of memory effects, $\Gamma(t)=2\delta(t)/\tau_m$ and $G(t)$ instantaneously reaches $G(\infty)$.
	For short memory times ($\tau_i \ll \tau_m$), $G(t)$ plateaus to $G(\infty)$ at $t<\tau_m$, such that $\tau_p=\tau_m$.
    However, for long memory times ($\tau_i\geq\tau_m$) the accumulation of friction is slow and less dissipation occurs at short times, resulting in an increased persistence time, $\tau_p$.
	As such, $\tau_p>\tau_m$ indicates the presence of memory effects with memory contributions satisfying $\tau_i\geq\tau_m$.
    All memory effects that contribute to parts of the total friction accumulating on times $t>\tau_m$, such as large values of $c$ and $d$, prolong the ballistic regime in the MSD, as seen in Figs.~\ref{fig_msd_free_alpha_n5}d-f.
	
	For certain combinations of $c$ and $d$, we see that oscillations appear in the MSD.
    The amplitude and duration of these oscillations increase with increasing $d/c$ when $c>1$, as can be seen when comparing Figs.~\ref{fig_msd_free_alpha_n5}h,i.
    When $c < 1$ and $d > 1$, or $c > 1$ and $d < 1$, and the first memory time is of the same order as the inertial time $\tau_m$, the form of Eq.~\eqref{eq_cd} causes the kernel $\Gamma(t)$ to decay on the scale of $\tau_m$.
    As a result, the kernel integral $G(t)$ saturates at $t \approx \tau_m$.
    In this case, a direct transition from the ballistic to the diffusive regime is observed in the MSD, which resembles the memoryless case, as in Fig.~\ref{fig_msd_free_alpha_n5}d.
	For increasing $c$, we also observe an increasingly long intermediate subdiffusive regime in the MSD up to the longest memory time $\tau_n=\tau_1 c^{n-1}$, as shown in Figs.~\ref{fig_msd_free_alpha_n5}a-c for $\tau_1=\tau_m$.
	The time-dependent exponent $\alpha(t)$ in Fig.~\ref{fig_msd_free_alpha_n5}i slightly oscillates, reflecting the individual exponential memory components, while its mean for $\tau_1<t<\tau_n$ is still given by Eq. \eqref{eq_alpha_cd_main}.
    In the SI Sec. \ref{sec_subdiff_c_n}, we show that for particularly large values of $c$, the minima in $\alpha(t)$ separate and additional intermediate diffusive regimes with $\alpha=1$ emerge.

	\begin{figure}
		\hspace{-5mm}
		\includegraphics{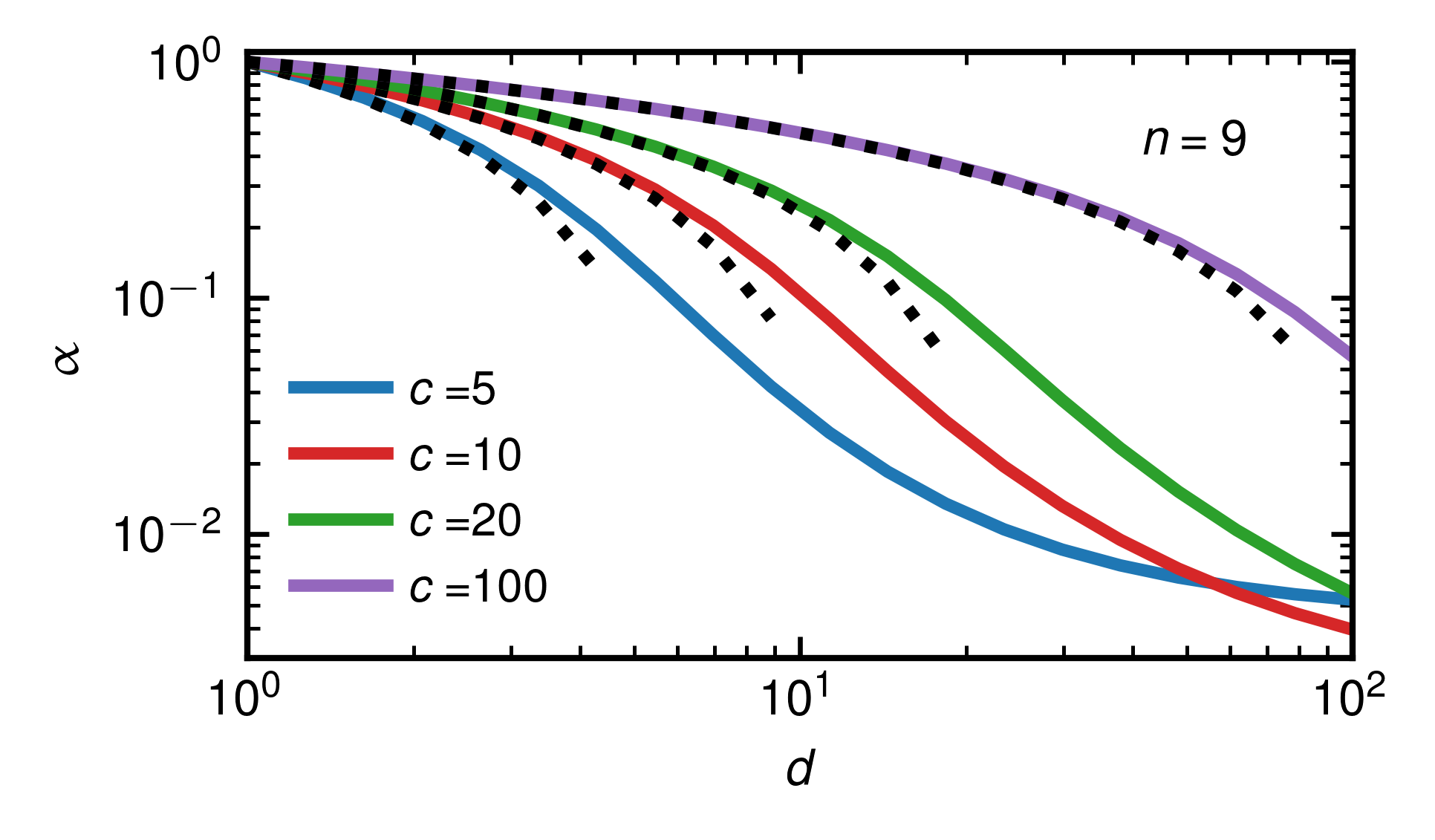}
		\caption{Exponent $\alpha$, obtained by fitting $\Gamma(t)\propto t^{-\alpha}$ to the multi-exponential kernel Eq.~\eqref{eq_kern_expo} with $n=9$ (solid lines), is compared to the prediction of Eq.~\eqref{eq_alpha_cd_main} (dotted lines) for different $c$ and $d$.}
		\label{fig_alpha_dn9}
	\end{figure}
    
    In Fig.~\ref{fig_alpha_dn9}, we compare the prediction of Eq.~\eqref{eq_alpha_cd_main} (dotted lines) to the scaling exponents $\alpha$ obtained by fitting power laws $\propto t^{-\alpha}$ to multi-exponential kernels $\Gamma(t)$ Eq.~\eqref{eq_kern_expo} (solid lines) over a finite range of time.
    The fitting details and examples for different values of $n$ are given in SI Sec. \ref{sec_subdiff_c_n}.
	For different $c$ and $d$, Eq.~\eqref{eq_alpha_cd_main} is seen to be a good description of the subdiffusive scaling exponent, as long as $c>d$.
	This scaling of the friction kernel translates into the intermediate subdiffusive scaling of the MSD for $\tau_p<t<\tau_n$, provided that $\alpha<1$ \cite{burov_critical_2008}, as shown in Figs.~\ref{fig_msd_free_alpha_n5}c,f.

	\subsection{Diffusion in a harmonic potential}\label{Sec_Harm}
\begin{figure}
		\centering
		\includegraphics{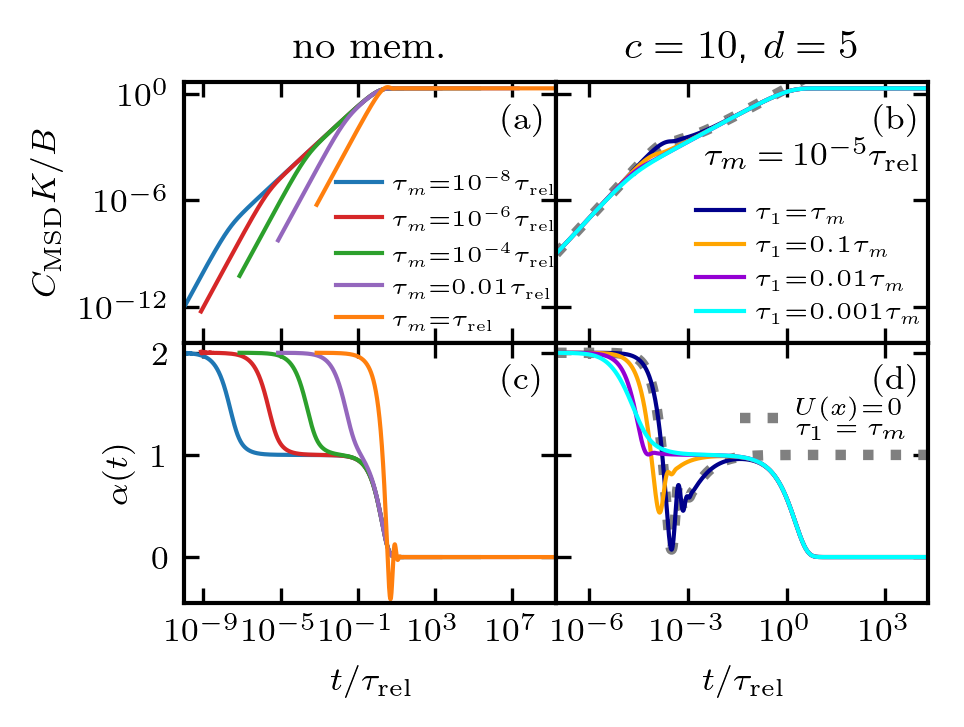}
		\hspace{-5mm}
		\caption{Analytical results for the MSD in harmonic confinement.
		The upper row depicts MSDs and the lower row the respective time-dependent exponent $\alpha(t)$ for (a),(c) without memory, governed by Eq.~\eqref{eq_le_wom}, and (b),(d) with $n=3$ exponential memory components, described by Eq.~\eqref{eq_kern_expo}.
		The gray dotted lines in (b) and (d) represent the result in the absence of a potential ($K=0$) with the other parameters equivalent to the dark-blue line.
		}
		\label{fig_msd_harmpot}
	\end{figure}

    Unlike the MSDs for freely-diffusing observables, which tend to pure diffusion ($\alpha=1$) in the long-time limit,
    the MSD for motion in a confining potential tends to a constant plateau value given by $C_{\rm{MSD}}(\infty)=2\langle x^2 \rangle$.
    Indeed, our analytic result for the MSD in Eq.~\eqref{eq_msd} for diffusion in a harmonic potential $U(x)=Kx^2/2$ (Figs.~\ref{fig_msd_harmpot}a and b) reach a finite plateau value with corresponding $\alpha(t)$ that tend to zero in the long-time limit, as shown in Figs.~\ref{fig_msd_harmpot}c and d.
	
	In Fig.~\ref{fig_msd_harmpot}a, we see that the MSD for the memoryless case, described by Eq.~\eqref{eq_le_wom}, transitions from ballistic to diffusive behavior before reaching the long-time plateau.
    This occurs if the inertial time $\tau_m$ is small compared to the relaxation time in the harmonic potential given by
	\begin{equation}
		\label{eq_tau_rel}
		\tau_{\rm{rel}} = 1/(\tau_m K)\,.
	\end{equation}
    Eq.~\eqref{eq_tau_rel} defines the damping time of a harmonic oscillator with oscillation frequency $K^{1/2}$.
    It is the time scale at which the confinement effects due to the potential become important for the MSD.
	For systems with a persistence time shorter than the theoretical oscillation period, $\tau_p<K^{-1/2}$, no oscillations arise from interactions with the potential.
    In this case, all features in the MSD that occur on timescales shorter than $\tau_{\rm{rel}}$ are not influenced by the potential.
    In the memoryless limit, where $\tau_p = \tau_m$, oscillations in the MSD - and consequently in $\alpha(t)$ - due to interactions with the potential occur only when $\tau_m$ is close to or larger than $\tau_{\rm{rel}}$.
    Such oscillations can be seen in Fig.~\ref{fig_msd_harmpot}c for the orange line, where $\tau_m=\tau_{\rm{rel}}=K^{-1/2}$.
	The motion in a harmonic potential with memory exhibits the same features in the MSD as for the free case on time scales smaller than the relaxation time $\tau_{\rm{rel}}$, as seen by comparing the dark-blue line with the gray dotted line in Figs.~\ref{fig_msd_harmpot}b,d.
	Furthermore, the data in Figs.~\ref{fig_msd_harmpot}b and d exhibit the limit of the standard Langevin dynamics when all memory times are shorter than the inertial time, $\tau_i\ll \tau_m$, leading to a direct transition from $\alpha=2$ to $\alpha=1$ at $t=\tau_m$ for the cyan line.
	
	\subsection{Memoryless diffusion in the presence of an energy barrier}
\begin{figure*}
		\centering
        \includegraphics[width=\textwidth]{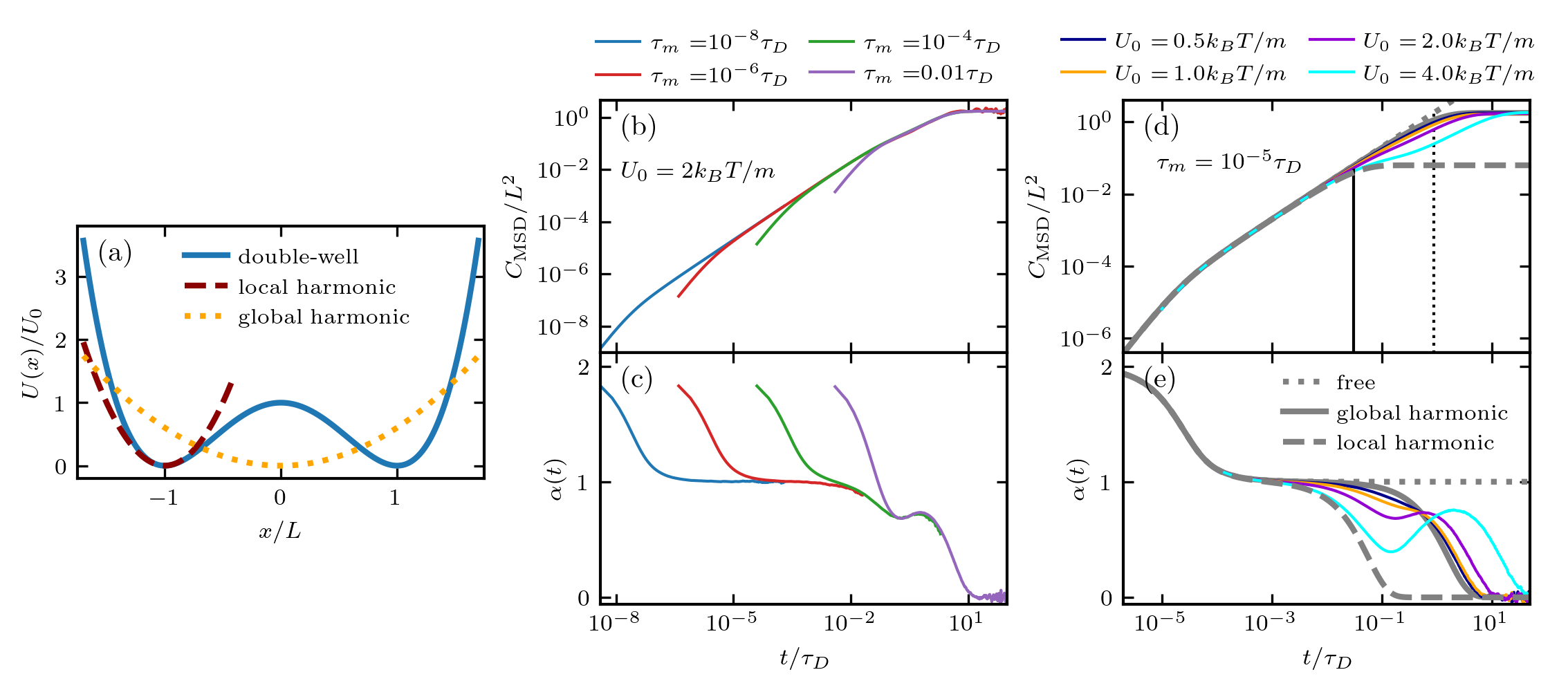}
		\caption{(a) Double-well potential (Eq.~\eqref{eq_double_well_def}, blue solid line) with corresponding local harmonic approximation around $x/L=-1$ (brown dashed line).
        The global harmonic potential (orange dotted line) leads to the same positional variance $\langle x^2\rangle$ as the double-well potential.
        (b)-(e) MSDs and time-dependent exponents $\alpha(t)$ for simulations in a double-well potential without memory;
        (b) and (c) results for different inertial times $\tau_m$ with fixed barrier height $U_0=2k_BT/m$.
        (d) and (e) Results for different barrier heights $U_0$ with fixed $\tau_m=10^{-5}\tau_D$.
        The dotted gray curves in (d) and (e) represent the analytical result for $U(x)=0$ with the same memory kernel parameters as used in the simulations.
        The solid gray curves represent the analytical result in the global harmonic potential, and the dashed gray curves represent the analytical result in the local harmonic potential, both for $U_0=4B$.
        The solid vertical line in (d) shows $\tau_{\rm{rel}}^{\rm{loc}}$ (Eq.~\eqref{eq_tau_loc}), and
        the vertical dotted line shows $\tau^{\rm{glob*}}_{\rm{rel}}$ (Eq.~\eqref{eq_tau_rel_mem}), both for $U_0=4B$.
        }
		\label{fig_dw_msd_nomem}
	\end{figure*}

    Systems with distinct meta-stable states, e.g. proteins, chemical reactants, or particles moving across coexisting phases, exhibit energy barriers, which can lead to subdiffusion \cite{evers_colloids_2013, goychuk_finite-range_2020}.
    However, as we demonstrated in the previous sections, memory-dependent friction by itself can lead to subdiffusion, even in the absence of a potential or energy barrier.
    To distinguish barrier from memory effects, we introduce the double-well potential
	\begin{equation}
		\label{eq_double_well_def}
		U(x) = U_0 \left( \left(\frac{x}{L}\right)^2 - 1 \right)^2\,,
	\end{equation}
	with barrier height $U_0$, which is a simplified, symmetric analog of the potential for the $\alpha_3$D protein shown in Fig.~\ref{fig1_motivation}a.
    Since it is analytically intractable to derive the MSD for the dynamics of a GLE in a non-harmonic potential, we turn to simulations \cite{simulation_git2025}.
	Here, we introduce the diffusion time $\tau_D=L^2/(B\tau_m)$, which is the average time to diffuse over a characteristic length $L$ in the absence of a potential.
	In the case of the double-well potential, we set $L$ as the distance between the minimum and the barrier position, as shown in Fig.~\ref{fig_dw_msd_nomem}a.

    In Figs.~\ref{fig_dw_msd_nomem}b-e, we discuss the effect of energy barriers on the MSD for a memoryless system. 
    In Figs.~\ref{fig_dw_msd_nomem}b and c, we show that for small inertial times $\tau_m \ll \tau_D$, there are extended regions of free diffusion in the MSD.
    That is, given that $\tau_p<K_{\rm{loc}}^{-1/2}$, which assures the absence of harmonic oscillations that might otherwise arise due to interaction with the potential.
    For times greater than the local harmonic relaxation time, defined by
	\begin{equation}
		\label{eq_tau_loc}
		\tau_{\rm{rel}}^{\rm{loc}} = 1/(\tau_m K_{\rm{loc}})\,,
	\end{equation}
    the observable interacts locally with the potential well, where $K_{\rm{loc}}$ is the single-well harmonic strength of the well, given by $K_{\rm{loc}}=8U_0/L^2$ (Fig.~\ref{fig_dw_msd_nomem}a).
    The system behaves much like the harmonically confined system discussed in the previous section, as seen by comparing Figs.~\ref{fig_msd_harmpot}a,c to Figs.~\ref{fig_dw_msd_nomem}b,c.
	However, in the case of the double-well potential, we observe the onset of additional subdiffusive behavior for $t>\tau_{\rm{rel}}^{\rm{loc}}$, characterized by non-monotonic behavior of $\alpha(t)$ (Fig.~\ref{fig_dw_msd_nomem}c), which, as discussed previously \cite{evers_colloids_2013, goychuk_finite-range_2020}, is entirely due to the barrier-crossing dynamics.
    For barrier heights $U_0<2B$, the MSD is well captured by the analytical result in a harmonic potential with a coupling strength $K_{\rm{glob}}$, chosen such that $B/K_{\rm{glob}}=\langle x^2 \rangle$, where we compute the variance, $\langle x^2 \rangle \propto \int_{-\infty}^{\infty} x^2 e^{-U(x)/B} dx$, numerically (Fig.~\ref{fig_dw_msd_nomem}d,e).
    $K_{\rm{glob}}$ is the effective global harmonic coupling strength associated with the quartic double-well potential. As such, $\tau_{\rm{rel}}^{\rm{glob}}$ is the global relaxation time obtained by using $K=K_{\rm{glob}}$ in Eq.~\eqref{eq_tau_rel}.
    The agreement between the harmonic prediction with $K_{\rm{glob}}$ and the MSD extracted from the double-well simulations for sufficiently low barriers ($U_0<2B$) indicates that the main effect of the double-well potential is the confinement at long times.
    In fact, for such small barriers, we find that the simulation MSDs agree well with the analytical result in the absence of an external potential up to the relaxation time of the global harmonic potential $\tau_{\rm{rel}}^{\rm{glob}}$ (vertical dotted line in Fig.~\ref{fig_dw_msd_nomem}d).
    This further emphasizes that the MSD is insensitive to sufficiently low barriers.
    It should be noted that such low barriers are observed in fast folding proteins \cite{dalton_fast_2023} or small polypeptides \cite{ayaz_non-markovian_2021}.
    However, these systems are known to be strongly non-Markovian.
    As we will see in the next section, memory has no influence on this insensitivity to low barriers.
    
    As we see in Fig.~\ref{fig_dw_msd_nomem}a, the minimum of a local well in the double-well potential can be approximated by a local harmonic potential. 
    Therefore, we can describe the short-time dynamics by our analytical results in a harmonic potential, which is suitable for times less than the relaxation time in the local well $\tau_{\rm{rel}}^{\rm{loc}}$, which we show as a vertical black line in Fig.~\ref{fig_dw_msd_nomem}d for $U_0=4B$.
    From the simulation results in Figs.~\ref{fig_dw_msd_nomem}d,e, we see that, for short times ($t<\tau_{\rm{rel}}^{\rm{loc}}$), the MSD in a double-well potential does indeed follow the prediction of the local harmonic well.
    The dip in the time-dependent exponent $\alpha(t)$ for $t>\tau_{\rm{rel}}^{\rm{loc}}$, observed in Fig.~\ref{fig_dw_msd_nomem}e, becomes more pronounced with increasing barrier height.
    It originates from the combined effect of local confinement at short times followed by barrier crossing.
    Therefore, potential energy landscapes with sufficiently high energy barriers can induce sub-diffusive behavior in the MSD, even in the absence of memory effects.
    
    \subsection{Diffusion in the presence of memory and energy barriers}
    \begin{figure*}
		\centering
        \includegraphics[width=\textwidth]{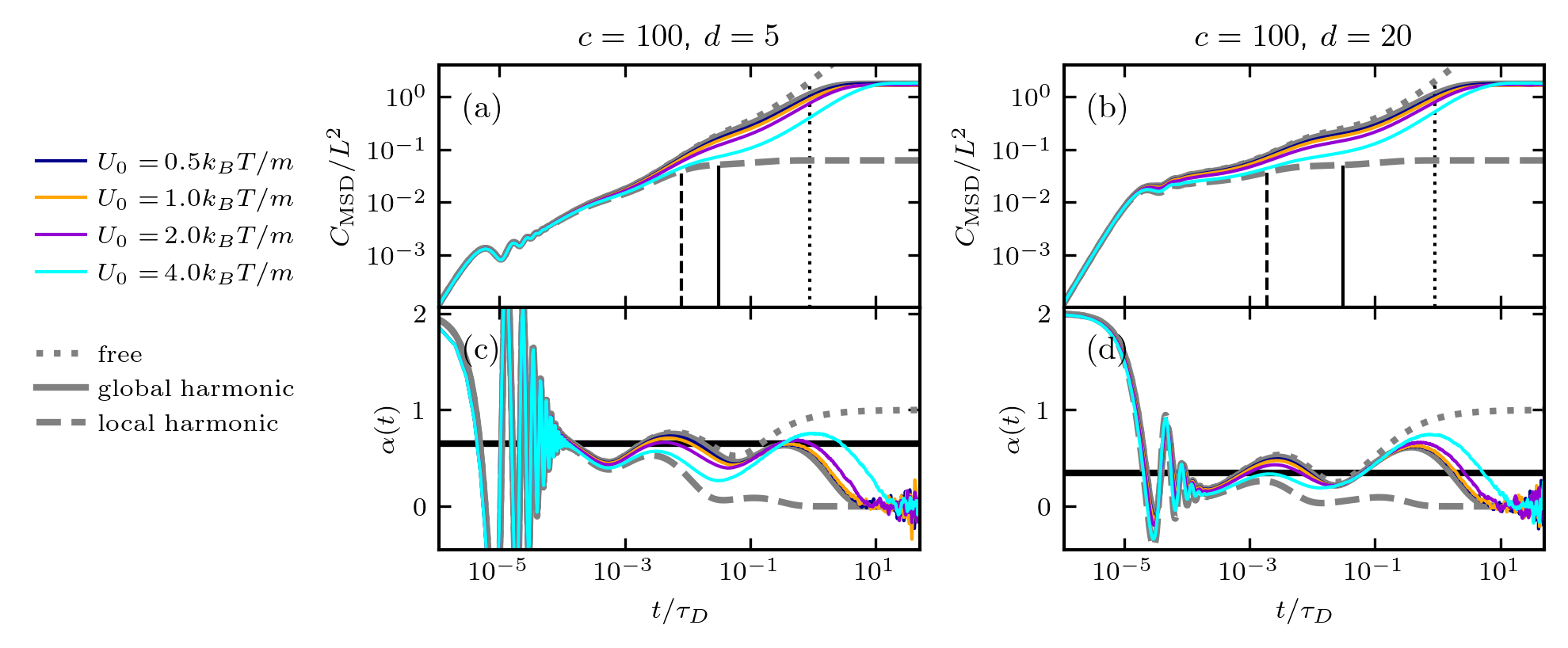}
		\caption{MSDs and time-dependent exponents $\alpha(t)$ from GLE simulations in a double-well potential with multi-exponential memory ($n=3$) over a range of potential energy barrier heights $U_0$.
        The inertial time is $\tau_m=10^{-8}\tau_D$, and the shortest memory time is $\tau_1=10^3\tau_m$.
        (a) and (c) Memory time ratio $c=100$ and friction amplitude ratio $d=5$, as defined in Eq.~\eqref{eq_cd}. 
        (b) and (d) $c=100$ and $d=20$.
        The dotted gray curves represent the analytical result for $U(x)=0$ with the same memory kernel parameters as used in the simulations.
        The solid gray curves represent the analytical result in the global harmonic potential, and the dashed gray curves represent the analytical result in the local harmonic potential, both for $U_0=4B$.
        The solid vertical lines show $\tau_{\rm{rel}}^{\rm{loc}}$ (Eq.~\eqref{eq_tau_loc}), the dashed vertical lines show $\tau_{\rm{rel}}^{\rm{loc*}}$ (Eq.~\eqref{eq_tau_loc_mem}), and the dotted vertical lines show $\tau^{\rm{glob*}}_{\rm{rel}}$ (Eq.~\eqref{eq_tau_rel_mem}), all for $U_0=4B$.
        }
		\label{fig_dw_msd_mem}
	\end{figure*}
    We now compare subdiffusion that is generated by memory effects to barrier-induced subdiffusion in a double-well potential.
    Simulations in the presence of memory are performed using Markovian embedding techniques \cite{simulation_git2025, brunig_time-dependent_2022,kappler2018memory}.
    In Figs.~\ref{fig_dw_msd_mem}a and b, we show MSDs for systems with memory kernels characterized by two different combinations of $c$ and $d$, as defined by Eq.~\ref{eq_cd}.
    We introduce a local harmonic relaxation time for systems with memory, $\tau_{\rm{rel}}^{\rm{loc*}}$, which is the timescale at which the friction force is balanced by the local harmonic restoration force, given by the self-consistent equation
    \begin{equation}
        \label{eq_tau_loc_mem}
        \tau_{\rm{rel}}^{\rm{loc*}} = G(\tau_{\rm{rel}}^{\rm{loc*}}) / K_{\rm{loc}}\,,
    \end{equation}
    which is determined numerically using an iteration procedure, similar to the persistence time $\tau_p$ defined in Eq.~\ref{eq_def_taup}.
    This prediction of $\tau_{\rm{rel}}^{\rm{loc*}}$, shown as dashed vertical lines for $U_0=4B$ in Figs.~\ref{fig_dw_msd_mem}a and b, agrees with the time at which the MSD of the double-well simulation deviates from the analytical result in the local harmonic well.
    $\tau_{\rm{rel}}^{\rm{loc*}}$ is smaller than the corresponding $\tau_{\rm{rel}}^{\rm{loc}}$, because the latter incorporates the total friction integral $G(\infty)$, while the former depends on an incomplete accumulation of friction, since $\Gamma(t)$ acts over a range of time scales.
    When friction is dominated by contributions with long memory times, which can be achieved by increasing $d$ for fixed $c$ (see Fig.~\ref{fig_msd_free_alpha_n5}), faster processes experience less friction.
    This has the effect of reducing $\tau_{\rm{rel}}^{\rm{loc*}}$, which can be seen by comparing the positions of the vertical dashed lines in Figs.~\ref{fig_dw_msd_mem}a and b. 
    
    Similarly, we introduce the relaxation time in the global potential for systems with memory-dependent friction, $\tau^{\rm{glob*}}_{\rm{rel}}$, which marks the deviation between the freely diffusing MSD and the MSD in the global harmonic well
	\begin{equation}
		\label{eq_tau_rel_mem}
		\tau^{\rm{glob*}}_{\rm{rel}} = G(\tau^{\rm{glob*}}_{\rm{rel}}) / K_{\rm{glob}}\,.
	\end{equation}
    Eq.~\ref{eq_tau_rel_mem} is analogous to Eq.~\eqref{eq_tau_rel} for systems with memory.
    In Figs.~\ref{fig_dw_msd_mem}a and b, we see that $\tau^{\rm{glob*}}_{\rm{rel}}$, indicated with dotted vertical lines, aligns precisely with the deviation times between the free diffusion MSD and the MSDs in the global harmonic potential.
    Much like in the case of local relaxation, where $\tau_{\rm{rel}}^{\rm{loc*}} <\tau_{\rm{rel}}^{\rm{loc}}$, the analog for the global relaxation is $\tau_{\rm{rel}}^{\rm{glob*}}<\tau_{\rm{rel}}^{\rm{glob}}$, i.e. the relaxation times in the presence of memory are always smaller than the relaxation times in the corresponding Markovian limits.
    These conditions are satisfied for any monotonically decreasing friction kernel.
    For low barriers (up to heights $U_0=2B$), we see in Figs.~\ref{fig_dw_msd_mem}c and d that the time-dependent exponent $\alpha(t)$ is well described by the analytical free-diffusion prediction (dotted lines) for $t<\tau^{\rm{glob*}}_{\rm{rel}}$.
    This means that for low enough barriers, the dynamics are dominated by the friction memory for $t<\tau^{\rm{glob*}}_{\rm{rel}}$, and that the energy barriers have negligible influence on the system dynamics.
    In general, the MSD and the corresponding $\alpha(t)$ in Fig.~\ref{fig_dw_msd_mem} for a particular memory kernel $\Gamma(t)$ are always bounded between the limiting free-diffusion behavior and the local harmonic approximation.
    The shape of the local potential becomes increasingly influential for large barrier heights.
    We observe that the subdiffusive contribution to the MSDs for $t<\tau_{\rm{rel}}^{\rm{loc*}}$ exhibits an exponent $\alpha$ close to the prediction by Eq.~\eqref{eq_alpha_cd_main} (solid horizontal black line in Figs.~\ref{fig_dw_msd_mem}c and d), regardless of barrier height.
    Therefore, even for systems with large barriers, memory effects still contribute significantly to the overall subdiffusivity.

    Taken together, we can summarize the main implications of Figs.~\ref{fig_dw_msd_nomem} and \ref{fig_dw_msd_mem} as follows:
    For small inertial times ($\tau_m\ll\tau_D$) and low energy barriers ($U_0 < 2B$), the dynamics are dominated by memory effects until $t=\tau^{\rm{glob*}}_{\rm{rel}}$, after which the global confinement dominates.
    For higher barrier heights, the dynamics exhibit memory-induced subdiffusivity until $t=\tau_{\rm{rel}}^{\rm{loc*}}$, after which the subdiffusive dynamics are due to a combination of memory effects and the shape of the potential.
    Clearly, the presence of energy barriers leads to additional subdiffusive behavior, in line with previous studies \cite{evers_colloids_2013,goychuk_finite-range_2020}.
    However, our findings suggest that, especially for small barrier heights and highly damped systems, the dynamics are dominated by memory effects, in line with recent findings for protein folding \cite{dalton_fast_2023, dalton2024ph}.
	\section{Conclusions}
    We analyze the dynamics of an observable experiencing multi-exponential friction memory in the framework of the generalized Langevin equation in the presence and absence of potential energy profiles $U(x)$.
	We find that, in the absence of a potential, $U(x)=0$, exponential memory components with exponentially spaced memory times and amplitudes produce subdiffusive behavior with the characterizing MSD exponent $\alpha$ predicted by Eq.~\eqref{eq_alpha_cd_main}.
	The short-time ballistic regime of the MSD, characterized by $\alpha=2$, is prolonged by memory effects and its termination is captured by the persistence time $\tau_p$, which is a generalization of the inertial time $\tau_m$ for systems with memory and defined in Eq.~\eqref{eq_def_taup}.
	In the presence of energy barrier heights $U_0\lesssim 2B$ (corresponding to $U_0\lesssim 2k_BT/m$), dynamics are dominated by memory effects for times smaller than the global harmonic relaxation time $t<\tau^{\rm{glob*}}_{\rm{rel}}$, defined in Eq.~\eqref{eq_tau_rel_mem}, and are well described by analytical results in an appropriately chosen global harmonic potential.
	In particular, the subdiffusive scaling $\alpha$ is predicted by the same Eq.~\eqref{eq_alpha_cd_main} as in the case without external potential, highlighting the importance of memory effects compared to effects due to the potential.
    In the case of strong damping $\tau_m\ll\tau_D$, the global harmonic relaxation time $\tau^{\rm{glob*}}_{\rm{rel}}$ becomes large such that the dynamics are governed by memory effects for long times.
    Additionally, we mention that the estimation of $\alpha$ via Eq.~\eqref{eq_alpha_cd_main} allows for the estimation of friction components and memory times on time scales below the temporal resolution of data by fitting the MSD behavior with parameters $c$ and $d$, as defined in Eq.~\eqref{eq_cd}.
	
	Effects of the potential landscape on the MSD become important for barrier heights above $U_0\sim 2B$.
	Similarly to previous findings \cite{evers_colloids_2013,goychuk_fingerprints_2021}, we observe a subdiffusive regime for sufficiently high energy barriers.
	Nevertheless, the dynamics in local minima are well described by our analytical results up to the local relaxation time $\tau_{\rm{rel}}^{\rm{loc*}}$, defined in Eq.~\eqref{eq_tau_loc_mem}, which depends on both the local shape of the potential and the memory, as demonstrated by simulations in a double-well potential.
	For shorter times $t<\tau_{\rm{rel}}^{\rm{loc*}}$, memory effects dominate the dynamics and after that, memory effects still influence the MSD significantly.

    Overall, we demonstrate that in order to accurately describe dynamics in subdiffusive systems, one must correctly account for both the potential energy landscape and memory effects.
    By disentangling contributions to subdiffusion by the potential landscape and memory effects, we confirm that, in many instances, the energy barriers are less influential than the friction in determining the system dynamics, as was shown recently to be the case for fast-folding proteins \cite{dalton_fast_2023}.
    However, our analysis is general and our findings are expected to have implications beyond the example of protein folding, shedding light on the origin of subdiffusive dynamics in a wide range of physical, chemical, and biological systems.

    \section*{Acknowledgements}
    We acknowledge funding by the Deutsche Forschungsgemeinschaft (DFG) through grant CRC 1449 “Dynamic Hydrogels at Biointerfaces”, Project ID 431232613, Project A03.

    \section*{Data availability}
    The code to simulate the GLE and implementations of the MSD formulas for multi-exponential memory are available at \url{https://github.com/kanton42/msd_subdiffusion}
    
	\bibliography{msd_theo}
	
	\makeatletter\@input{aux_file_si.tex}\makeatother
\end{document}


\maketitle

    \section{Determining MSD exponent $\alpha$ from multiexponential memory}\label{sec_derivation_alpha}
	Starting from the friction kernel expression given in the main text Eq.~\eqref{eq_kern_expo} with memory-time scaling $c$ and friction-coefficient scaling $d$ according to Eq.~\eqref{eq_cd}, one obtains the values of the kernel $\Gamma(t_i)=\Gamma_i$ at times $t_i = \tau_1 c^{i-1}$ approximately as
	\begin{equation}
		\label{eq_kern_discrete}
		\Gamma_i = e^{-1}\frac{\gamma_1}{\tau_1} \sum_{j=i}^{n} \frac{d^{j-1}}{c^{j-1}}\,,
	\end{equation}
	where contributions $j<i$ are neglected, as they are exponentially small for $c>1$.
	We define the ratio of the scaling factors $b=d/c$ and split the sum in Eq.~\eqref{eq_kern_discrete}, which leads to
	\begin{align}
		\Gamma_i &= e^{-1} \frac{\gamma_1}{\tau_1} \left( \sum_{j=i}^{\infty}b^{j-1} - \sum_{j=n+1}^{\infty}b^{j-1} \right)\\
		&= e^{-1} \frac{\gamma_1}{\tau_1} \left( b^{i-1}\sum_{k=0}^{\infty}b^{k} - b^n \sum_{k=0}^{\infty}b^{k} \right)\,.
	\end{align}
	In the case of $b<1$, i.e. $c>d$, the geometric sum converges to $\sum_{k=0}^{\infty}b^{k} = 1/(1-b)$.
    Thus the kernel at time $t_i$ is given by
	\begin{equation}
		\label{eq_gammai_discrete_res}
		\Gamma_i = e^{-1} \frac{\gamma_1}{\tau_1} \frac{b^{i-1} - b^n}{1-b}\,.
	\end{equation}
	The logarithm of Eq.~\eqref{eq_gammai_discrete_res} reads
	\begin{equation}
		\ln (\Gamma_i) = 1 + \ln \left(\frac{\gamma_1}{\tau_1}\right) + (i-1) \ln (b) + \ln(1-b^{n-i+1}) - \ln (1-b)\,,
	\end{equation}
	which leads to the logarithmic derivative between $t_i$ and $t_{i+1}$ as
	\begin{equation}
		\label{eq_kern_log_deriv}
		\frac{\rm{d} \ln (\Gamma(t))}{\rm{d}\ln (t)}|_{t=t_i} = \dfrac{\ln (\Gamma_{i+1}) - \ln (\Gamma_{i})}{\ln (t_{i+1}) - \ln(t_i)} = \frac{\ln (b) + \ln (1-b^{n-i}) - \ln (1-b^{n-i+1})}{\ln (c)}\,.
	\end{equation}
	In the case of $n\gg i-1$, the logarithmic derivative Eq.~\eqref{eq_kern_log_deriv} simplifies to
	\begin{equation}
		\label{eq_log_der_approx}
		\frac{\rm{d} \ln (\Gamma(t))}{\rm{d}\ln (t)} = \frac{\ln (b)}{\ln (c)} =  \frac{\ln (d/c)}{\ln (c)}\,.
	\end{equation}
	This logarithmic derivative is the exponent of the friction kernel $\Gamma(t)\propto t^{-\alpha}$, as shown in Fig.~\ref{fig_msd_free_alpha_n5} in the main text. Since a scaling of the friction kernel $\Gamma(t)\propto t^{-\alpha}$ leads to $C_{\rm{MSD}}(t)\propto t^\alpha$, for $\alpha<1$ \cite{burov_critical_2008}, we can estimate the scaling exponent of the MSD via Eq.~\eqref{eq_log_der_approx} as a function of $c$ and $d$ as
	\begin{equation}
		\label{eq_alpha_cd}
		\alpha(c, d) = \frac{\ln (c/d)}{\ln (c)}\,.
	\end{equation}
    %
	\begin{figure*}
		\centering
		\includegraphics[width=\textwidth]{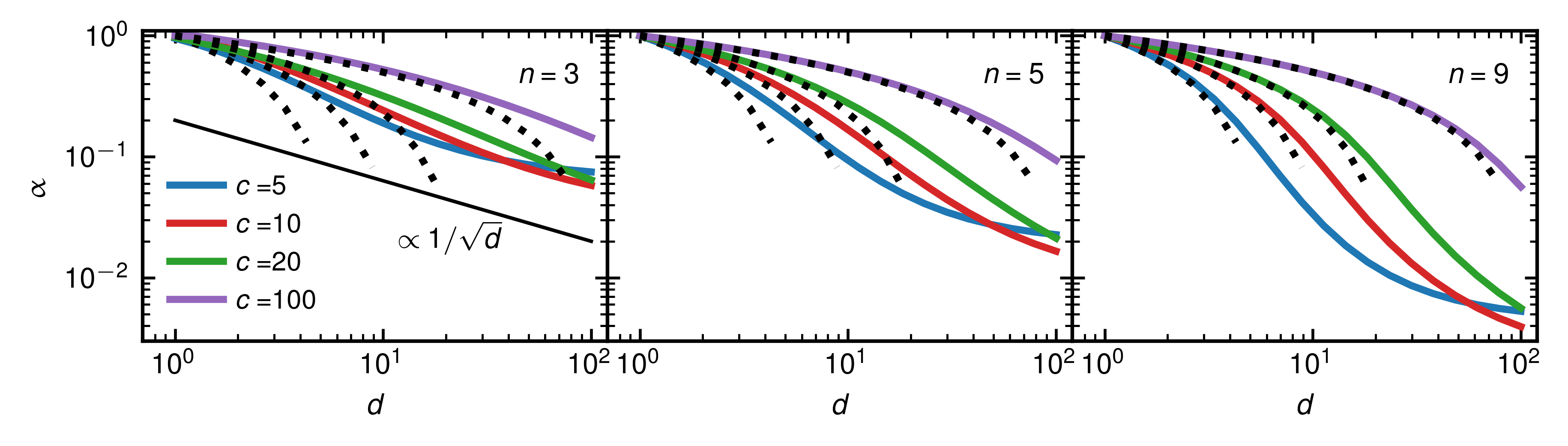}
		\caption{The behavior of $\alpha$ obtained from fits of $\Gamma(t)\propto t^{-\alpha}$ to the multiexponential kernel Eq.~\eqref{eq_kern_expo} is compared to the prediction by Eq.~\eqref{eq_alpha_cd} (dotted lines) for different $c$ and $d$ and for $n=3$, $n=5$ and $n=9$.}
		\label{fig_alpha_cdn}
	\end{figure*}

	In Fig.~\ref{fig_alpha_cdn} we compare the result of Eq.~\eqref{eq_alpha_cd} to the $\alpha$ values obtained from double-logarithmic fits of $\Gamma (t)\propto t^{-\alpha}$ to the friction kernel Eq.~\eqref{eq_kern_expo} between $\tau_1<t<\tau_n$ for different values of $c$, $d$, and $n$.
    The fits minimize the squared difference of the logarithms on a logarithmically spaced time axis between $\tau_1<t<\tau_n$ for $n=9$ exponentials.
	As expected from the derivation, Eq.~\eqref{eq_alpha_cd} describes the scaling exponent $\alpha$ most accurately for high numbers of exponential kernel contributions $n$, shown by the good agreement between fitted $\alpha$ and Eq.~\eqref{eq_alpha_cd} for $n=9$ in Fig.~\ref{fig_alpha_cdn}c.
	But even for $n=3$ and $n=5$ the scaling exponent of the friction kernel $\Gamma(t)$ is well predicted for $c>d$, where the deviation of the prediction Eq.~\eqref{eq_alpha_cd} to the observed scaling becomes larger with increasing $d$, as expected by the assumption $c>d$ in the derivation.
	
	In contrast to the subdiffusive regimes originating from positive exponential friction components, it was shown before that negative exponential friction components lead to additional intermediate ballistic regimes \cite{mitterwallner_negative_2020}.
	
	In the long time limit, $\alpha$ always tends towards the long time diffusive value of $\alpha=1$, and thus, the estimation of the scaling exponent by Eq.~\eqref{eq_alpha_cd} is not applicable for long times $t>\tau_n$.
    Clearly, this is expected since exponential memory kernels involve a longest time scale, in contrast to power-law memory kernels, which lead to subdiffusive behavior on all time scales when not using short-time and long-time cutoffs.
    Examples of the transition to the long-time diffusive behavior at $t=\tau_n$ are shown in Figs.~\ref{fig_msd_free_alpha_n3}d,h.
	
	\section{Derivation of MSD for the GLE with multi-exponential memory}\label{sec_msd_expo_derivation}
	In Fourier space, the GLE (Eq.~\eqref{eq_gle_wom}) can be written as
	\begin{equation}
		\label{eq_ft_gle}
		\tilde{x}(\omega) = \tilde{\chi}(\omega)\tilde{F}_R(\omega) \,,
	\end{equation}
	where $\tilde{\chi}(\omega)$ is the Fourier transform of the position response function, which for a harmonic potential $U(x)=Kx^2/2$ takes the form
	\begin{equation}
		\label{eq_def_response_fun}
		\tilde{\chi}(\omega) = \left(K -\omega^2 + i\omega\tilde{\Gamma}_+(\omega) \right)^{-1}
	\end{equation}
	with $\tilde{\Gamma}_+(\omega)$ being the half-sided Fourier transform of the memory kernel, as defined via
	\begin{equation}
		\label{eq_def_halfside_kernel_transform}
		\tilde{\Gamma}_+(\omega) = \int_{0}^{\infty}e^{-i\omega t} \Gamma(t) dt \,.
	\end{equation}
	The Fourier transform of the position correlation function $C_{xx}(t) = \langle x(0) x(t)\rangle $ can be written as
	\begin{equation}
		\label{eq_ft_pos_corr}
		\tilde{C}_{xx}(\omega) = B\tilde{\chi}(\omega) \tilde{\Gamma}(\omega) \tilde{\chi}(-\omega) \,,
	\end{equation}
	where we made use of the relation between the second moment of the random force and $\Gamma(t)$, Eq.~\eqref{eq_fdt}, and the mean squared velocity $B=\langle \dot{x}^2 \rangle$.
    Equation \eqref{eq_ft_pos_corr} can be rewritten as
	\begin{equation}
		\tilde{C}_{xx}(\omega) = -\frac{B}{i\omega} \left( \tilde{\chi}(\omega) - \tilde{\chi}(-\omega) \right) \,,
	\end{equation}
	which leads to the MSD given as
	\begin{eqnarray}
		C_{\rm{MSD}}(t) &=& 2 (C_{xx}(0) - C_{xx}(t)) \nonumber\\
		&=& B \int_{-\infty}^{\infty} \frac{d\omega}{\pi} \frac{e^{i\omega t} - 1}{i\omega}\left( \tilde{\chi}(\omega) - \tilde{\chi}(-\omega) \right) \label{eq_msd_integral} \,.
	\end{eqnarray}
    %
	Generally, the response function for a sum of $n$ exponentials kernel
	\begin{equation}
		\label{eq_kernel_general_sum_expos}
		\Gamma(t)=\sum_{i}^{n}a_i e^{-t/\tau_i}
	\end{equation}
	can be written for a harmonic potential as
	\begin{equation}
		\label{eq_response_general_expos}
		\tilde{\chi}(\omega) = \left(K -\omega^2 + \sum_{j=1}^{n}\frac{i \omega a_j}{i\omega +\frac{1}{\tau_j}}\right)^{-1}\,,
	\end{equation}
	where we defined $a_i = \frac{\gamma_i}{\tau_i}$.
	Inserting Eq.~\eqref{eq_response_general_expos} into the MSD integral, Eq.~\eqref{eq_msd_integral}, one obtains
	\begin{equation}
		C_{\rm{MSD}}(t)=B\int_{-\infty}^{\infty}\frac{d\omega }{\pi} \frac{-2 p_1 (e^{i\omega t}-1) \sum_{j}^n \frac{a_j}{\tau_j} p_2(m\neq j)}{p_1p_2(K^2-2K\omega^2 +\omega^4) + 2\omega^2 (K - \omega^2) p_1 \sum_{j}^{n}a_j p_2(j\neq m) + p_2\sum_{j,k}^{n}\omega^2a_ja_k p_1(l,m\neq j,k)}\,,
		\label{eq_msd_harm_general_expos}
	\end{equation}
	with 
	\begin{equation}
		\label{eq_prod_p1}
		p_1 = \prod_{l, m}^{n}\left(\omega^2+\frac{i\omega}{\tau_m}-\frac{i\omega}{\tau_l}+\frac{1}{\tau_l\tau_m}\right)
	\end{equation}
	and
	\begin{equation}
		\label{eq_prod_p2}
		p_2 = \prod_{m=1}^{n}\left(\omega^2+\frac{1}{\tau_m^2}\right)\,.
	\end{equation}
	$p_1(l,m\neq j,k)$ denotes $p_1$ leaving out the indices $l=j$ and $m=k$ in the multiplication Eq.~\eqref{eq_prod_p1} and similarly, $p_2(m\neq j)$ denotes $p_2$ without the inclusion of the term with $m=j$ in Eq.~\eqref{eq_prod_p2}.
	From Eq.~\eqref{eq_msd_harm_general_expos}, one can bring the MSD into the form
	\begin{equation}
		C_{\rm{MSD}}(t) = B \int_{-\infty}^{\infty}\frac{d\omega}{\pi}\frac{\sum_{i}^{2n-1} k_i \omega^{2(i-1)}}{c_{n+2}\prod_{i=0}^{n+2}(\omega^2-\omega^2_i)}
		\label{eq_msd_polynomial_form}
	\end{equation}
	with $\omega^2_i$ being the solutions to the $n+2$ order polynomial in $\omega^2$
	\begin{equation}
		\label{eq_polynomial_general_sum_expos_harm}
		\sum_{i=0}^{n+2}c_i \omega^{2i} = 0\,.
	\end{equation}
	The constants $c_i$ and $k_i$ can be read off by comparing Eqs.~\eqref{eq_msd_polynomial_form} and \eqref{eq_msd_harm_general_expos}, where $c_i$ are not to be confused with the memory time ratio $c$.
	The MSD takes the form
	\begin{equation}
		\label{msd_final_res_general_sum_expos_harm}
		C_{\rm{MSD}}(t) = \frac{B}{c_{n+2}} \left( \sum_{i=1}^{n+2}\frac{e^{-\sqrt{-\omega^2_i} t} - 1}{\sqrt{-\omega^2_i} \prod_{j\neq i} (\omega^2_i-\omega^2_j) }  \sum_{m=1}^{2n-1}k_m \omega_i^{2m-2}  \right)\,,
	\end{equation}
	
	In the case $K=0$, Eq.~\eqref{eq_msd_harm_general_expos} simplifies to
	\begin{equation}
		\label{eq_msd_general_expos}
		C_{\rm{MSD}}(t)=B\int_{-\infty}^{\infty}\frac{d\omega }{\pi} \frac{-2 p_1 (e^{i\omega t}-1) \sum_{j}^n \frac{a_j}{\tau_j}\prod_{j\neq m}(\omega^2+\frac{1}{\tau_m^2})}{\omega^4 p_1 \left[p_2 - 2\sum_{j}^{n}a_j \prod_{j\neq m}(\omega^2+\frac{1}{\tau_m^2})\right] + p_2\sum_{j,k}^{n}\omega^2a_ja_k\prod_{l,m\neq j,k}(\omega^2 + \frac{i\omega}{\tau_l} - \frac{i\omega}{\tau_m} + \frac{1}{\tau_l\tau_m})}\,.
	\end{equation}
	and the MSD takes the form
	\begin{equation}
		C_{\rm{MSD}}(t) = B \int_{-\infty}^{\infty}\frac{d\omega}{\pi}\frac{\sum_{i}^{2n-1} k_i \omega^{2(i-1)}}{c_{n+2}\prod_{i=1}^{n+2}(\omega^2-\omega^2_i)},
		\label{eq_msd_polynomial_form_free}
	\end{equation}
	with $\omega^2_i$ now given by the roots of the $n+1$ order polynomial
	\begin{equation}
		\label{eq_polynomial_general_sum_expos}
		\sum_{i=1}^{n+2}c_i \omega^{2i-2} = 0\,.
	\end{equation}
	The MSD for $K=0$ thus always takes the form
	\begin{equation}
		\label{msd_final_res_general_sum_expos}
		C_{\rm{MSD}}(t) = \frac{B}{c_{n+2}} \left( \frac{(-1)^{n }k_1 t}{\prod_{j=1}^{n}\omega^2_j} + \sum_{i=1}^{n+1}\frac{e^{-\sqrt{-\omega_i^2} t} - 1}{\sqrt{-\omega_i^2} \prod_{j\neq i} (\omega_i^2-\omega^2_j) }  \sum_{m=1}^{2n-1}k_m \omega_i^{2m-4}  \right)\,.
	\end{equation}
	Here, the first term proportional to $t$ determines the long time diffusivity, since the other terms decay exponentially as a function of time.
	As an example, we now consider the case of $K=0$ and $n=3$ in the memory kernel 
    Eq.~\eqref{eq_kernel_general_sum_expos}.
    From Eq.~\eqref{eq_msd_general_expos} one obtains
	\begin{equation}
		\label{eq_msd_int_inserted_triexp}
		C_{\rm{MSD}}(t) = B \int_{-\infty}^{\infty} \frac{d\omega}{\pi} \frac{(e^{i\omega t} - 1) (k_1 + k_2\omega^2 + k_3\omega^4)}{\omega^2(c_1 + c_2 \omega^2 + c_3 \omega^4 +c_4\omega^6 +c_5\omega^8)} \,.
	\end{equation}
	where the constants $c_i$ and $k_i$ are given as
	
	\begin{empheq}[box=\colorboxed{black}]{align}
		k_1 =& -2 (a_1 \tau_1 + a_2 \tau_2 + a_3 \tau_3) \nonumber \\
		k_2 =& -2 (a_2 \tau_1^2 \tau_2 + a_1 \tau_1 \tau_2^2 + 
		a_3 \tau_1^2 \tau_3 + a_3 \tau_2^2 \tau_3 + a_1 \tau_1 \tau_3^2 + 
		a_2 \tau_2 \tau_3^2) \nonumber \\
		k_3 =& -2 \tau_1 \tau_2 \tau_3 (a_3 \tau_1 \tau_2 + 
		a_2 \tau_1 \tau_3 + a_1 \tau_2 \tau_3) \nonumber \\
		c_1 =& (a_1 \tau_1 + a_2 \tau_2 + a_3 \tau_3)^2 \nonumber \\
		c_2 =& 1 - 2 a_2 \tau_2^2 + \tau_1^2 (-2 a_1 + (a_1 + a_2)^2 \tau_2^2) + 
		2 a_3 \tau_1 \tau_2 (a_2 \tau_1 + 
		a_1 \tau_2) \tau_3 + ((a_1 \tau_1 + a_2 \tau_2)^2 + 
		a_3^2 (\tau_1^2 + \tau_2^2)  \\
		+& 2 a_3 (-1 + a_1 \tau_1^2 + a_2 \tau_2^2)) \tau_3^2 \nonumber \\
		c_3 =& \tau_3^2 + \tau_2^2 (1 - 2 (a_2 + a_3) \tau_3^2) + \tau_1^2 (1 - 
		2 (a_1 + a_2) \tau_2^2 + (-2 (a_1 + a_3) + (a_1 + a_2 + 
		a_3)^2 \tau_2^2) \tau_3^2) \nonumber \\
		c_4 =& -\tau_2^2 \tau_3^2 + \tau_1^2 (-\tau_3^2 + \tau_2^2 (-1 + 
		2 (a_1 + a_2 + a_3) \tau_3^2)) \nonumber \\
		c_5 =& \tau_1^2 \tau_2^2 \tau_3^2 \nonumber
	\end{empheq}
	
	Regarding the integral in Eq.~\eqref{eq_msd_int_inserted_triexp} as a sum of the three terms containing $k_i$, we see that all terms have the same poles, where the term containing $k_1$ has the additional double pole at $\omega=0$. The solutions to
	\begin{equation}
		\label{eq_poles_msd}
		c_1 + c_2 \omega^2 + c_3 \omega^4 +c_4\omega^6 +c_5\omega^8 = 0 \,,
	\end{equation}
	which we denote by $\omega^2_i$, define the remaining poles of the integral Eq.~\eqref{eq_msd_int_inserted_triexp} as $\pm \sqrt{\omega^2_i}$. We rewrite the polynomial Eq.~\eqref{eq_poles_msd} using the roots $\omega^2_i$ as
	\begin{equation}
		c_1 + c_2 \omega^2 + c_3 \omega^4 +c_4\omega^6 +c_5\omega^8 = c_5 (\omega^2-\omega^2_1)(\omega^2-\omega^2_2)(\omega^2-\omega^2_3)(\omega^2-\omega^2_4)\,.
	\end{equation}
	
	Next we use the partial fraction decompositions
	\begin{eqnarray}
		\frac{1}{\omega^2\prod_{i=1}^{k}(\omega^2-\omega^2_i)} &=& \frac{(-1)^k}{\omega^2\prod_{i=1}^{k}\omega^2_i} + \sum_{i=1}^{k} \frac{1}{\omega^2_i (\omega^2-\omega^2_i) \prod_{j\neq i}(\omega^2_i-\omega^2_j)}\label{eq_partial_fraction_decomp1}
		\\
		\frac{\omega^{2n}}{\prod_{i=1}^{k}(\omega^2-\omega^2_i)} &=&  \sum_{i=1}^{k} \frac{\omega^{2n}_i}{(\omega^2-\omega^2_i)\prod_{j\neq i}(\omega^2_i-\omega^2_j)}\label{eq_partial_fraction_decomp3}
	\end{eqnarray}
	with $n\geq0$ to rewrite the fraction of Eq.~\eqref{eq_msd_int_inserted_triexp} as a sum of terms proportional to $(\omega^2-\omega^2_i)^{-1}$ and one term proportional to $\omega^{-2}$.
    Using the solutions of the integrals
	\begin{eqnarray}
		\label{eq_integrals_roots}
		\int_{-\infty}^{\infty}\frac{e^{i\omega t} -1}{\omega^2-\omega^2_i}d\omega  &=& \frac{\pi (e^{-\sqrt{-\omega^2_i}t} - 1)}{\sqrt{-\omega^2_i}}\\
		\int_{-\infty}^{\infty}\frac{ e^{i\omega t} - 1}{\omega^2}d\omega &=& -\pi t
	\end{eqnarray}
	for $t>0$ with the condition $\text{Re}(\omega^2_i)<0 \lor \text{Im}(\omega^2_i)\neq0 \land \text{Im}(\sqrt{\omega^2_i})\neq0$, we can rewrite the integral of the MSD Eq.~\eqref{eq_msd_int_inserted_triexp} as
	\begin{equation}
		\label{eq_msd_final_res_triexp}
		\colorboxed{black}{
			C_{\rm{MSD}}(t) = \frac{B}{c_5} \left( \frac{-k_1 t}{\omega^2_1\omega^2_2\omega^2_3\omega^2_4} + \sum_{i=1}^{4}\frac{e^{-\sqrt{-\omega^2_i} t} - 1}{\sqrt{-\omega^2_i} \prod_{j\neq i} (\omega^2_i-\omega^2_j) } \left[ \frac{k_1}{\omega^2_i} + k_2 + k_3 \omega^2_i \right] \right)} \,.
	\end{equation}
	
    The poles, i.e. the roots of the polynomial Eq.~\eqref{eq_poles_msd}, can be determined analytically for polynomials up to fourth order, which one obtains for $n=3$ exponentials without potential or $n=2$ exponentials in a harmonic potential.
    For any larger $n$, the roots of the polynomials of Eq.~\eqref{eq_polynomial_general_sum_expos} or Eq.~\eqref{eq_polynomial_general_sum_expos_harm}, have to be solved numerically. We determine those roots using \textit{sympy} in Python.
	
	The constants $c_i$ and $k_i$ in Eqs.~\eqref{eq_msd_general_expos} and \eqref{eq_polynomial_general_sum_expos} that define the MSD via Eq.~\eqref{msd_final_res_general_sum_expos} for a $n=5$ exponential memory kernel (Eq.~\eqref{eq_kernel_general_sum_expos}) without any potential are given by
    \begin{empheq}[box=\colorboxed{violet}]{align}
		c_5 =& \tau_3^2 \tau_4^2 \tau_5^2 + \tau_2^2 
		(\tau_3^2 \tau_4^2 + (\tau_3^2 + (1 - 
		2 (a_2 + a_3 + a_4 + 
		a_5) \tau_3^2) \tau_4^2) \tau_5^2) \nonumber\\
		+& \tau_1^2 [\tau_3^2 \tau_4^2 + (\tau_4^2 + \tau_3^2 (1 - 
		2 (a_1 + a_3 + a_4 + 
		a_5) \tau_4^2)) \tau_5^2 
		+ \tau_2^2 [\tau_4^2 + (1 - 2 (a_1 + a_2 + a_4 + a_5) \tau_4^2) \tau_5^2 \nonumber\\
		+& \tau_3^2 (1 - 2 (a_1 + a_2 + a_3 + 
		a_4) \tau_4^2 + (-2 (a_1 + a_2 + a_3 + a_5) + (a_1 + 
		a_2 + a_3 + a_4 + a_5)^2 \tau_4^2) \tau_5^2)]] \\
		c_6 =& \tau_2^2 \tau_3^2 \tau_4^2 \tau_5^2 + \tau_1^2 [\tau_3^2 \tau_4^2 \tau_5^2
		+ \tau_2^2 (\tau_3^2 \tau_4^2 + (\tau_3^2 + (1 - 
		2 (a_1 + a_2 + a_3 + a_4 + 
		a_5) \tau_3^2) \tau_4^2) \tau_5^2)] \nonumber\\
		c_7 =& \tau_1^2 \tau_2^2 \tau_3^2 \tau_4^2 \tau_5^2 \nonumber
	\end{empheq}
	
	\begin{empheq}[box=\colorboxed{violet}]{align}
		k_1 =& -2 (a_1 \tau_1 + a_2 \tau_2 + a_3 \tau_3 + a_4 \tau_4 + 
		a_5 \tau_5) \nonumber \\
		k_2 =& -2 (a_3 \tau_1^2 \tau_3 + a_3 \tau_2^2 \tau_3 + 
		a_4 \tau_1^2 \tau_4 + a_4 \tau_2^2 \tau_4
		+ a_4 \tau_3^2 \tau_4 + a_3 \tau_3 \tau_4^2 + 
		a_5 (\tau_1^2 + \tau_2^2 + \tau_3^2 + \tau_4^2) \tau_5 \nonumber \\
		+&(a_3 \tau_3 + a_4 \tau_4) \tau_5^2 
		+ a_2 \tau_2 (\tau_1^2 + \tau_3^2 + \tau_4^2 + \tau_5^2) + 
		a_1 \tau_1 (\tau_2^2 + \tau_3^2 + \tau_4^2 + \tau_5^2))\nonumber \\
		k_3 =& -2 [a_2 \tau_1^2 \tau_2 \tau_3^2 + 
		a_1 \tau_1 \tau_2^2 \tau_3^2 + 
		a_4 \tau_1^2 \tau_2^2 \tau_4
		+ a_4 \tau_1^2 \tau_3^2 \tau_4 + 
		a_4 \tau_2^2 \tau_3^2 \tau_4 + 
		a_2 \tau_1^2 \tau_2 \tau_4^2 
		+ a_1 \tau_1 \tau_2^2 \tau_4^2 + 
		a_1 \tau_1 \tau_3^2 \tau_4^2 \nonumber \\
		+& a_2 \tau_2 \tau_3^2 \tau_4^2 
		+ a_5 (\tau_3^2 \tau_4^2 + \tau_2^2 (\tau_3^2 + \tau_4^2) + 
		\tau_1^2 (\tau_2^2 + \tau_3^2 + \tau_4^2)) \tau_5 \nonumber \\
		+& (a_4 (\tau_1^2 + \tau_2^2 + \tau_3^2) \tau_4 + 
		a_2 \tau_2 (\tau_1^2 + \tau_3^2 + \tau_4^2) + 
		a_1 \tau_1 (\tau_2^2 + \tau_3^2 + \tau_4^2)) \tau_5^2 \nonumber \\
		+& a_3 \tau_3 (\tau_4^2 \tau_5^2 + \tau_2^2 (\tau_4^2 + 
		\tau_5^2) + \tau_1^2 (\tau_2^2 + \tau_4^2 + \tau_5^2))] \nonumber \\
		k_4 =& -2 (\tau_1 \tau_2 \tau_3 \tau_4 (a_4 \tau_1 \tau_2 
		\tau_3 + (a_3 \tau_1 \tau_2 + a_2 \tau_1 \tau_3 + 
		a_1 \tau_2 \tau_3) \tau_4)
		+ a_5 (\tau_2^2 \tau_3^2 \tau_4^2 + \tau_1^2 (\tau_3^2 
		\tau_4^2 + \tau_2^2 (\tau_3^2 + \tau_4^2))) \tau_5 \nonumber \\
		+& (\tau_1 
		\tau_2 \tau_3 (a_3 \tau_1 \tau_2 + a_2 \tau_1 \tau_3
		+ a_1 \tau_2 \tau_3) + 
		a_4 (\tau_2^2 \tau_3^2 + \tau_1^2 (\tau_2^2 + \
		\tau_3^2)) \tau_4 + (a_3 (\tau_1^2 + \tau_2^2) \tau_3 \nonumber \\
		+& a_2 \tau_2 (\tau_1^2 + \tau_3^2) + 
		a_1 \tau_1 (\tau_2^2 + \tau_3^2)) \tau_4^2) \tau_5^2) \nonumber \\
		k_5 =& -2 \tau_1 \tau_2 \tau_3 \tau_4 \tau_5 (a_5 \tau_1 
		\tau_2 \tau_3 \tau_4 + 
		a_4 \tau_1 \tau_2 \tau_3 \tau_5 + (a_3 \tau_1 \tau_2 + 
		a_2 \tau_1 \tau_3 + a_1 \tau_2 \tau_3) \tau_4 \tau_5) \nonumber \\
		c_1 =& (a_1 \tau_1 + a_2 \tau_2 + a_3 \tau_3 + a_4 \tau_4 + 
		a_5 \tau_5)^2 \nonumber \\
		c_ 2 =& 1 - 2 a_2 \tau_2^2 - 2 a_3 \tau_3^2 + 
		a_3^2 \tau_1^2 \tau_3^2 + a_3^2 \tau_2^2 \tau_3^2
		+ 2 a_3 a_4 \tau_1^2 \tau_3 \tau_4 + 
		2 a_3 a_4 \tau_2^2 \tau_3 \tau_4 - 2 a_4 \tau_4^2 + 
		a_4^2 \tau_1^2 \tau_4^2 + a_4^2 \tau_2^2 \tau_4^2 \nonumber \\
		+& a_3^2 \tau_3^2 \tau_4^2 
		+ 2 a_3 a_4 \tau_3^2 \tau_4^2 + 
		a_4^2 \tau_3^2 \tau_4^2 + 
		2 a_5 (a_4 (\tau_1^2 + \tau_2^2 + \tau_3^2) \tau_4
		+ a_3 \tau_3 (\tau_1^2 + \tau_2^2 + \tau_4^2)) \tau_5 \nonumber \\
		+& ((a_3 \tau_3 + a_4 \tau_4)^2 + 
		a_5^2 (\tau_1^2 + \tau_2^2 + \tau_3^2 + \tau_4^2) 
		+ 2 a_5 (-1 + a_3 \tau_3^2 + a_4 \tau_4^2)) \tau_5^2 + a_2^2 \tau_2^2 (\tau_1^2 + \tau_3^2 + \tau_4^2 + \tau_5^2) \nonumber \\
		+& a_1^2 \tau_1^2 (\tau_2^2 + \tau_3^2 + \tau_4^2 + 
		\tau_5^2) + 
		2 a_2 \tau_2 (a_5 \tau_5 (\tau_1^2 + \tau_3^2 + \tau_4^2
		+ \tau_2 \tau_5)
		+ a_4 \tau_4 (\tau_1^2 + \tau_3^2 + \tau_2 \tau_4 + 
		\tau_5^2) \nonumber \\
		+& a_3 \tau_3 (\tau_1^2 + \tau_2 \tau_3 + \tau_4^2 + 
		\tau_5^2)) 
		+ 2 a_1 \tau_1 (a_2 \tau_2 \tau_3^2 + a_4 \tau_2^2 \tau_4 + 
		a_4 \tau_3^2 \tau_4 + a_2 \tau_2 \tau_4^2 + 
		a_5 (\tau_2^2 + \tau_3^2 + \tau_4^2) \tau_5 \nonumber \\
		+& (a_2 \tau_2 + a_4 \tau_4) \tau_5^2 + 
		a_3 \tau_3 (\tau_2^2 + \tau_4^2 + \tau_5^2) + \tau_1 
		(-1 + a_2 \tau_2^2 + a_3 \tau_3^2 + a_4 \tau_4^2 + a_5 \tau_5^2))\nonumber \\
		c_3 =& \tau_4^2 + \tau_3^2 (1 - 
		2 (a_3 + a_4) \tau_4^2)
		+ \tau_5^2 + (-2 (a_3 + a_5) \tau_3^2 + (-2 (a_4 + a_5) + (a_3 + a_4 + 
		a_5)^2 \tau_3^2) \tau_4^2) \tau_5^2 \label{eq_constants_c_n5} \\ 
		+& 2 a_2 \tau_2 \tau_3 \tau_4 \tau_5 (a_5 \tau_3 \tau_4 + 
		a_4 \tau_3 \tau_5 + a_3 \tau_4 \tau_5)
		+ 2 a_1 \tau_1 [\tau_2 \tau_3 \tau_4 (a_4 \tau_2 \tau_3 + 
		a_3 \tau_2 \tau_4 + a_2 \tau_3 \tau_4) \nonumber \\
		+& a_5 (\tau_3^2 \tau_4^2 + \tau_2^2 (\tau_3^2 + 
		\tau_4^2)) \tau_5 
		+ (a_4 (\tau_2^2 + \tau_3^2) \tau_4 + a_3 \tau_3 (\tau_2^2 + \tau_4^2) + 
		a_2 \tau_2 (\tau_3^2 + \tau_4^2)) \tau_5^2] \nonumber \\
		+& \tau_2^2 [1 - 
		2 a_4 \tau_4^2 + \tau_3^2 (-2 a_3 + (a_3 + 
		a_4)^2 \tau_4^2) + 
		2 a_5 \tau_3 \tau_4 (a_4 \tau_3 + 
		a_3 \tau_4) \tau_5 \nonumber \\
		+& ((a_3 \tau_3 + a_4 \tau_4)^2 + 
		a_5^2 (\tau_3^2 + \tau_4^2) + 
		2 a_5 (-1 + a_3 \tau_3^2 + a_4 \tau_4^2)) \tau_5^2
		+ a_2^2 (\tau_4^2 \tau_5^2 + \tau_3^2 (\tau_4^2 + 
		\tau_5^2)) \nonumber \\
		+& 2 a_2 (-\tau_4^2 + (-1 + (a_4 + 
		a_5) \tau_4^2) \tau_5^2 + \tau_3^2 (-1 + (a_3 + 
		a_4) \tau_4^2 + (a_3 + 
		a_5) \tau_5^2))] \nonumber \\
		+& \tau_1^2 [1 - 2 a_3 \tau_3^2 +
		a_3^2 \tau_2^2 \tau_3^2 + 
		2 a_3 a_4 \tau_2^2 \tau_3 \tau_4 - 2 a_4 \tau_4^2
		+ a_4^2 \tau_2^2 \tau_4^2 + a_3^2 \tau_3^2 \tau_4^2 + 
		2 a_3 a_4 \tau_3^2 \tau_4^2 + a_4^2 \tau_3^2 \tau_4^2 \nonumber \\
		+& 2 a_5 (a_4 (\tau_2^2 + \tau_3^2) \tau_4 + 
		a_3 \tau_3 (\tau_2^2 + \tau_4^2)) \tau_5 \nonumber \\
		+& ((a_3 \tau_3 + a_4 \tau_4)^2 + a_5^2 (\tau_2^2 + \tau_3^2 + \tau_4^2) + 
		2 a_5 (-1 + a_3 \tau_3^2 + a_4 \tau_4^2)) \tau_5^2
		+ a_2^2 \tau_2^2 (\tau_3^2 + \tau_4^2 + \tau_5^2) \nonumber \\
		+& a_1^2 (\tau_4^2 \tau_5^2 + \tau_3^2 (\tau_4^2 + 
		\tau_5^2) + \tau_2^2 (\tau_3^2 + \tau_4^2 + \tau_5^2)) 
		+ 2 a_2 \tau_2 (\tau_3 \tau_4 (a_4 \tau_3 + a_3 \tau_4) +
		a_5 (\tau_3^2 + \tau_4^2) \tau_5 \nonumber \\
		+& (a_3 \tau_3 + a_4 \tau_4) \tau_5^2 + \tau_2 (-1 + 
		a_3 \tau_3^2 + a_4 \tau_4^2 + a_5 \tau_5^2))
		+ 2 a_1 (-\tau_4^2 + (-1 + (a_4 + 
		a_5) \tau_4^2) \tau_5^2 \nonumber \\
		+& \tau_3^2 (-1 + (a_3 + a_4) \tau_4^2 + (a_3 + 
		a_5) \tau_5^2) 
		+ \tau_2^2 (-1 + a_3 \tau_3^2 + 
		a_4 \tau_4^2 + a_5 \tau_5^2 + 
		a_2 (\tau_3^2 + \tau_4^2 + \tau_5^2)))]\nonumber \\
		c_4 =& \tau_3^2 \tau_4^2 + (\tau_4^2 + \tau_3^2 (1 - 
		2 (a_3 + a_4 + a_5) \tau_4^2)) \tau_5^2 
		+ 2 a_1 \tau_1 \tau_2 \tau_3 \tau_4 \tau_5 (a_5 \tau_2 
		\tau_3 \tau_4 \\
        +& (a_4 \tau_2 \tau_3 + a_3 \tau_2 \tau_4 + 
		a_2 \tau_3 \tau_4) \tau_5) \nonumber \\
		+& \tau_2^2 (\tau_5^2 \
		+ \tau_4^2 (1 - 2 (a_2 + a_4 + a_5) \tau_5^2) + \tau_3^2 (1 - 
		2 (a_2 + a_3 + 
		a_4) \tau_4^2 \nonumber \\
		+& (-2 (a_2 + a_3 + a_5) + (a_2 + a_3 + 
		a_4 + a_5)^2 \tau_4^2) \tau_5^2))
		+ \tau_1^2 [\tau_3^2 + \tau_4^2 - 
		2 (a_1 + a_3 + a_4) \tau_3^2 \tau_4^2 \nonumber \\
		+& (1 - 2 (a_1 + a_3 + 
		a_5) \tau_3^2 + (-2 (a_1 + a_4 + a_5) + (a_1 + a_3 + 
		a_4 + a_5)^2 \tau_3^2) \tau_4^2) \tau_5^2 \nonumber \\
		+& 2 a_2 \tau_2 \tau_3 \tau_4 \tau_5 (a_5 \tau_3 
		\tau_4 + a_4 \tau_3 \tau_5 + a_3 \tau_4 \tau_5)
		+ \tau_2^2 [1 - 2 a_4 \tau_4^2 + \tau_3^2 (-2 a_3 + (a_3 + 
		a_4)^2 \tau_4^2) \nonumber \\
		+& 2 a_5 \tau_3 \tau_4 (a_4 \tau_3 + 
		a_3 \tau_4) \tau_5 
		+ ((a_3 \tau_3 + a_4 \tau_4)^2 + a_5^2 (\tau_3^2 + \tau_4^2) + 
		2 a_5 (-1 + a_3 \tau_3^2 + 
		a_4 \tau_4^2)) \tau_5^2 \nonumber \\
		+& a_1^2 (\tau_4^2 \tau_5^2 + \tau_3^2 (\tau_4^2 + 
		\tau_5^2))
		+ a_2^2 (\tau_4^2 \tau_5^2 + \tau_3^2 (\tau_4^2 + 
		\tau_5^2)) \nonumber \\
		+& 2 a_2 (-\tau_4^2 + (-1 + (a_4 + a_5) \tau_4^2) \tau_5^2 + \tau_3^2 (-1 + 
		(a_3 + a_4) \tau_4^2 + (a_3 + a_5) \tau_5^2)) \nonumber \\
		+& 2 a_1 (-\tau_4^2 + (-1 + (a_2 + a_4 + 
		a_5) \tau_4^2) \tau_5^2 + \tau_3^2 (-1 + (a_2 + 
		a_3 + a_4) \tau_4^2 + (a_2 + a_3 + a_5) \tau_5^2))]] \nonumber \,.
	\end{empheq}
	For $n=3$ exponentials in a harmonic potential, the constants of Eqs.~\eqref{eq_msd_harm_general_expos} and \eqref{eq_polynomial_general_sum_expos_harm} are given by
	\begin{empheq}[box=\colorboxed{black}]{align}
		k_1 =& -2 (a \tau_1 + b \tau_2 + c \tau_3) \nonumber \\
		k_2 =& -2 (b \tau_1^2 \tau_2 + a \tau_1 \tau_2^2 + 
		c \tau_1^2 \tau_3 + c \tau_2^2 \tau_3 + a \tau_1 \tau_3^2 + 
		b \tau_2 \tau_3^2) \nonumber \\
		k_3 =& -2 \tau_1 \tau_2 \tau_3 (c \tau_1 \tau_2 + 
		b \tau_1 \tau_3 + a \tau_2 \tau_3) \nonumber \\
		c_0 =& K^2 \nonumber \\
		c_1 =& (a \tau_1 + b \tau_2 + c \tau_3)^2 + 
		K^2 (\tau_1^2 + \tau_2^2 + \tau_3^2) + 
		2 K (-1 + a \tau_1^2 + b \tau_2^2 + c \tau_3^2) \nonumber \\
		c_2 =& 1 - 2 K \tau_1^2 - 2 b \tau_2^2 - 2 K \tau_2^2 + 
		b^2 \tau_1^2 \tau_2^2 + 2 b K \tau_1^2 \tau_2^2 + 
		K^2 \tau_1^2 \tau_2^2 + 
		2 b c \tau_1^2 \tau_2 \tau_3 \label{eq_constants_ck_n3_harm} \\
		+& (-2 c - 2 K + c^2 \tau_1^2 + 
		2 c K \tau_1^2 + 
		K^2 \tau_1^2 + (b + c + K)^2 \tau_2^2) \tau_3^2 \nonumber \\
		+& a^2 \tau_1^2 (\tau_2^2 + \tau_3^2) + 
		2 a \tau_1 (\tau_1 (-1 + (b + K) \tau_2^2) + 
		c \tau_2^2 \tau_3 + ((c + K) \tau_1 + b \tau_2) \tau_3^2) \nonumber \\
		c_3 =& \tau_3^2 + \tau_2^2 (1 - 
		2 (b + c + K) \tau_3^2) + \tau_1^2 (1 - 
		2 (a + b + 
		K) \tau_2^2 \nonumber \\
		+& (-2 (a + c + K) + (a + b + c + 
		K)^2 \tau_2^2) \tau_3^2) \nonumber \\
		c_4 =& \tau_2^2 \tau_3^2 + \tau_1^2 (\tau_3^2 + \tau_2^2 (1 - 
		2 (a + b + c + K) \tau_3^2)) \nonumber \\
		c_5 =& \tau_1^2 \tau_2^2 \tau_3^2 \nonumber
	\end{empheq}
	For $n=5$ exponentials in a harmonic potential, the constants of Eqs.~\eqref{eq_msd_harm_general_expos} and \eqref{eq_polynomial_general_sum_expos_harm} are given by
	\begin{empheq}[box=\colorboxed{violet}]{align}
		c_4 =& \tau_3^2 \tau_4^2 + (\tau_4^2 + \tau_3^2 (1 - 
		2 (a_3 + a_4 + a_5 + K) \tau_4^2)) \tau_5^2 + 
		2a_1\tau_1 \tau_2 \tau_3 \tau_4 \tau_5 (a_5 \tau_2 \tau_3 \tau_4 \nonumber \\
        +& (a_4 \tau_2 \tau_3 + a_3 \tau_2 \tau_4 +
		a_2 \tau_3 \tau_4) \tau_5) \nonumber \\
		+& \tau_2^2 (\tau_4^2 + (1 - 
		2 (a_2 + a_4 + a_5 + K) \tau_4^2) \tau_5^2 + \tau_3^2 (1 - 
		2 (a_2 + a_3 + a_4 + 
		K) \tau_4^2 + (-2 (a_2 + a_3 + a_5 + K) \nonumber \\
		+& (a_2 + a_3 + a_4 + a_5 + 
		K)^2 \tau_4^2) \tau_5^2))\nonumber \\
		+& \tau_1^2 (\tau_3^2 + \tau_4^2 - 
		2(a_1+ a_3 + a_4 + K) \tau_3^2 \tau_4^2 + (1 - 
		2(a_1+ a_3 + a_5 + 
		K) \tau_3^2 \nonumber \\
		+& (-2(a_1+ a_4 + a_5 + K) +(a_1+ a_3 + a_4 + a_5 + 
		K)^2 \tau_3^2) \tau_4^2) \tau_5^2 \nonumber \\
		+& 2 a_2 \tau_2 \tau_3 \tau_4 \tau_5 (a_5 \tau_3 \tau_4 + 
		a_4 \tau_3 \tau_5 + a_3 \tau_4 \tau_5) \nonumber \\
		+& \tau_2^2 [1 - 
		2a_1\tau_3^2 - 2 a_2 \tau_3^2 - 2 a_3 \tau_3^2 - 
		2 K \tau_3^2 - 2a_1\tau_4^2 - 2 a_2 \tau_4^2 - 
		2 a_4 \tau_4^2 - 2 K \tau_4^2 + a_1^2 \tau_3^2 \tau_4^2
		+
		2a_1a_2 \tau_3^2 \tau_4^2 \nonumber \\
		+& a_2^2 \tau_3^2 \tau_4^2 + 
		2a_1a_3 \tau_3^2 \tau_4^2 + 2 a_2 a_3 \tau_3^2 \tau_4^2 + 
		a_3^2 \tau_3^2 \tau_4^2 + 2a_1a_4 \tau_3^2 \tau_4^2 \nonumber\\
		+&2 a_2 a_4 \tau_3^2 \tau_4^2 + 2 a_3 a_4 \tau_3^2 \tau_4^2 +
		a_4^2 \tau_3^2 \tau_4^2 + 2a_1K \tau_3^2 \tau_4^2 \nonumber \\
		+& 
		2 a_2 K \tau_3^2 \tau_4^2 + 2 a_3 K \tau_3^2 \tau_4^2 + 
		2 a_4 K \tau_3^2 \tau_4^2 + K^2 \tau_3^2 \tau_4^2 \label{eq_constants_c4to7_n5_harm} \\
		+& 
		2 a_5 \tau_3 \tau_4 (a_4 \tau_3 + 
		a_3 \tau_4) \tau_5 + (-2 (a_2 + a_5 + K) + (a_2 + a_3 + a_5 + 
		K)^2 \tau_3^2
		+
		2 a_3 a_4 \tau_3 \tau_4 \nonumber \\
		+& (a_2 + a_4 + a_5 + K)^2 \tau_4^2 + 
		a_1^2 (\tau_3^2 + \tau_4^2) \nonumber \\
        +& 2a_1(-1 + (a_2 + a_3 + a_5 + K) \tau_3^2 + (a_2 + a_4 + a_5 + 
		K) \tau_4^2)) \tau_5^2]) \nonumber \\
		c_5 =& \tau_3^2 \tau_4^2 \tau_5^2 + \tau_2^2 (\tau_3^2 \tau_4^2 + (\tau_3^2 + (1 - 
		2 (a_2 + a_3 + a_4 + a_5 + 
		K) \tau_3^2) \tau_4^2) \tau_5^2) \nonumber \\
		+& \tau_1^2 (\tau_3^2 \tau_4^2 + (\tau_4^2 + \tau_3^2 (1 - 
		2(a_1+ a_3 + a_4 + a_5 + 
		K) \tau_4^2)) \tau_5^2 \nonumber \\
        +& \tau_2^2 (\tau_4^2 + (1 - 
		2(a_1+ a_2 + a_4 + a_5 + 
		K) \tau_4^2) \tau_5^2 \nonumber \\
		+& \tau_3^2 (1 - 
		2(a_1+ a_2 + a_3 + a_4 + 
		K) \tau_4^2 + (-2(a_1+ a_2 + a_3 + a_5 + K) \nonumber \\
        +&(a_1+ a_2 + a_3 + 
		a_4 + a_5 + K)^2 \tau_4^2) \tau_5^2))) \nonumber \\
		c_6 =& \tau_2^2 \tau_3^2 \tau_4^2 \tau_5^2 + \tau_1^2 (\tau_3^2 \tau_4^2 \tau_5^2 + \tau_2^2 (\tau_3^2 \tau_4^2 + (\tau_3^2 + (1 - 2(a_1+ a_2 + a_3 + a_4 + a_5 + K) \tau_3^2) \tau_4^2) \tau_5^2)) \nonumber \\
		c_7 =& \tau_1^2 \tau_2^2 \tau_3^2 \tau_4^2 \tau_5^2 \nonumber
	\end{empheq}

	\begin{empheq}[box=\colorboxed{violet}]{align}
		k_1 =& -2 ( a_1\tau_1 +  a_2 \tau_2 +  a_3\tau_3 +  a_4\tau_4 +  a_5\tau_5) \nonumber \\
		k_2 =& -2 ( a_3\tau_1^2 \tau_3 +  a_3\tau_2^2 \tau_3 + 
		a_4\tau_1^2 \tau_4 +  a_4\tau_2^2 \tau_4 \nonumber \\
        +&  a_4\tau_3^2 \tau_4 + a_3\tau_3 \tau_4^2 + 
		a_5(\tau_1^2 + \tau_2^2 + \tau_3^2 + \tau_4^2) \tau_5 + ( a_3\tau_3 +  a_4\tau_4) \tau_5^2 \nonumber \\
		+&  a_2 \tau_2 (\tau_1^2 + \tau_3^2 + \tau_4^2 + \tau_5^2) + 
		a_1\tau_1 (\tau_2^2 + \tau_3^2 + \tau_4^2 + \tau_5^2)) \nonumber \\
		k_3 =& -2 ( a_2 \tau_1^2 \tau_2 \tau_3^2 + 
		a_1\tau_1 \tau_2^2 \tau_3^2 +  a_4\tau_1^2 \tau_2^2 \tau_4 + 
		a_4\tau_1^2 \tau_3^2 \tau_4 \nonumber \\
        +& a_4\tau_2^2 \tau_3^2 \tau_4 + 
		a_2 \tau_1^2 \tau_2 \tau_4^2 +  a_1\tau_1 \tau_2^2 \tau_4^2 +
		a_1\tau_1 \tau_3^2 \tau_4^2 +  a_2 \tau_2 \tau_3^2 \tau_4^2 \nonumber \\
		+& a_5(\tau_3^2 \tau_4^2 + \tau_2^2 (\tau_3^2 + \tau_4^2) + \tau_1^2 (\tau_2^2 + \tau_3^2 + \tau_4^2)) \tau_5 + ( a_4(\tau_1^2 + \tau_2^2 + \tau_3^2) \tau_4 + 
		a_2 \tau_2 (\tau_1^2 + \tau_3^2 + \tau_4^2) \nonumber \\
		+& a_1\tau_1 (\tau_2^2 + \tau_3^2 + \tau_4^2)) \tau_5^2 + 
		a_3\tau_3 (\tau_4^2 \tau_5^2 + \tau_2^2 (\tau_4^2 + \
		\tau_5^2) + \tau_1^2 (\tau_2^2 + \tau_4^2 + \tau_5^2))) \nonumber \\
		k_4 =& -2 \tau_1 \tau_2 \tau_3 \tau_4 ( a_4\tau_1 \tau_2 \tau_3 + ( a_3\tau_1 \tau_2 +  a_2 \tau_1 \tau_3 + 
		a_1\tau_2 \tau_3) \tau_4) - 
		2  a_5(\tau_2^2 \tau_3^2 \tau_4^2 + \tau_1^2 (\tau_3^2 \tau_4^2 + \tau_2^2 (\tau_3^2 + \tau_4^2))) \tau_5 \nonumber \\
		-& 
		2 (\tau_1 \tau_2 \tau_3 ( a_3\tau_1 \tau_2 +  a_2 \tau_1 \tau_3 + 
		a_1\tau_2 \tau_3) + 
		a_4(\tau_2^2 \tau_3^2 + \tau_1^2 (\tau_2^2 + \tau_3^2)) \tau_4 + ( a_3(\tau_1^2 + \tau_2^2) \tau_3 + 
		a_2 \tau_2 (\tau_1^2 + \tau_3^2) \nonumber \\
		+& 
		a_1\tau_1 (\tau_2^2 + \tau_3^2)) \tau_4^2) \tau_5^2 \nonumber \\
		k_5 =& -2 \tau_1 \tau_2 \tau_3 \tau_4 \tau_5 ( a_5\tau_1 \tau_2 \tau_3 \tau_4 +  a_4\tau_1 \tau_2 \tau_3 \tau_5 + 
		a_3\tau_1 \tau_2 \tau_4 \tau_5 + 
		a_2 \tau_1 \tau_3 \tau_4 \tau_5 + 
		a_1\tau_2 \tau_3 \tau_4 \tau_5) \nonumber \\
		c_0 =& K^2 \nonumber \\
		c_1 =& ( a_1\tau_1 +  a_2 \tau_2 +  a_3\tau_3 +  a_4\tau_4 +  a_5\tau_5)^2 \nonumber \\
		+& 
		K^2 (\tau_1^2 + \tau_2^2 + \tau_3^2 + \tau_4^2 + \tau_5^2) + 
		2 K (-1 +  a_1\tau_1^2 +  a_2 \tau_2^2 +  a_3\tau_3^2 +  a_4\tau_4^2 + 
		a_5\tau_5^2) \nonumber \\
		c_2 =& 1 - 2  a_2 \tau_2^2 +  a_2^2 \tau_1^2 \tau_2^2 + 
		2  a_2  a_3\tau_1^2 \tau_2 \tau_3 - 2  a_3\tau_3^2 + 
		a_3^2 \tau_1^2 \tau_3^2 +  a_2^2 \tau_2^2 \tau_3^2 + 
		2  a_2  a_3\tau_2^2 \tau_3^2 + a_3^2 \tau_2^2 \tau_3^2 \nonumber \\
		+& 
		2  a_2  a_4\tau_1^2 \tau_2 \tau_4 + 2  a_3 a_4\tau_1^2 \tau_3 \tau_4 + 
		2  a_3 a_4\tau_2^2 \tau_3 \tau_4 + 2  a_2  a_4\tau_2 \tau_3^2 \tau_4 
        - 2  a_4\tau_4^2 +  a_4^2 \tau_1^2 \tau_4^2 +  a_2^2 \tau_2^2 \tau_4^2 \nonumber \\
        +& 2  a_2  a_4\tau_2^2 \tau_4^2 +  a_4^2 \tau_2^2 \tau_4^2 
		+ 2  a_2  a_3\tau_2 \tau_3 \tau_4^2 + a_3^2 \tau_3^2 \tau_4^2 + 
		2  a_3 a_4\tau_3^2 \tau_4^2 +  a_4^2 \tau_3^2 \tau_4^2 + 
		2  a_5( a_4(\tau_1^2 + \tau_2^2 + \tau_3^2) \tau_4 \nonumber \\
        +& 
		a_3\tau_3 (\tau_1^2 + \tau_2^2 + \tau_4^2)
		+ a_2 \tau_2 (\tau_1^2 + \tau_3^2 + \tau_4^2)) \tau_5 + (( a_2 \tau_2 +  a_3\tau_3 +  a_4\tau_4)^2 \nonumber \\
        +& a_5^2 (\tau_1^2 + \tau_2^2 + \tau_3^2 + \tau_4^2) + 
		2  a_5(-1 +  a_2 \tau_2^2 +  a_3\tau_3^2 +  a_4\tau_4^2)) \tau_5^2
		+ a_1^2 \tau_1^2 (\tau_2^2 + \tau_3^2 + \tau_4^2 + \tau_5^2) + 
		K^2 (\tau_3^2 \tau_4^2 \nonumber \\
		+& (\tau_3^2 + \tau_4^2) \tau_5^2 + \tau_2^2 (\tau_3^2
		+ \tau_4^2 + \tau_5^2) + \tau_1^2 (\tau_2^2 + \tau_3^2 + \tau_4^2 + \tau_5^2))
		+
		2 K (-\tau_4^2 + \tau_3^2 (-1 + ( a_3+ 
		a_4) \tau_4^2) \nonumber \\
		-& \tau_5^2 + (( a_3+  a_5) \tau_3^2 + ( a_4+ 
		a_5) \tau_4^2) \tau_5^2 
		+ \tau_1^2 (-1 +  a_2 \tau_2^2 
		+
		a_3\tau_3^2 +  a_4\tau_4^2 +  a_5\tau_5^2) \label{eq_constants_c1to3_k_n5_harm} \\
		+& \tau_2^2 (-1 + 
		a_2 \tau_3^2 +  a_3\tau_3^2 +  a_2 \tau_4^2 + 
		a_4\tau_4^2 + ( a_2 +  a_5) \tau_5^2)) + 
		2  a_1\tau_1 ( a_2 \tau_2 \tau_3^2 +  a_4\tau_2^2 \tau_4
		+ a_4\tau_3^2 \tau_4 +  a_2 \tau_2 \tau_4^2 \nonumber \\
		+& 
		a_5(\tau_2^2 + \tau_3^2 + \tau_4^2) \tau_5 + ( a_2 \tau_2 + 
		a_4\tau_4) \tau_5^2 + 
		a_3\tau_3 (\tau_2^2 + \tau_4^2 + \tau_5^2) \nonumber \\
		+& \tau_1 (-1 + 
		a_2 \tau_2^2 +  a_3\tau_3^2 +  a_4\tau_4^2 +  a_5\tau_5^2 + 
		K (\tau_2^2 + \tau_3^2 + \tau_4^2 + \tau_5^2))) \nonumber \\
		c_3 =& \tau_4^2 + \tau_3^2 (1 - 
		2 ( a_3+  a_4+ K) \tau_4^2) + \tau_5^2 + (-2 ( a_3+  a_5+ 
		K) \tau_3^2 + (-2 ( a_4+  a_5+ K) \nonumber \\
		+& ( a_3+  a_4+  a_5+ 
		K)^2 \tau_3^2) \tau_4^2) \tau_5^2
		+
		2  a_2 \tau_2 \tau_3 \tau_4 \tau_5 ( a_5\tau_3 \tau_4 + 
		a_4\tau_3 \tau_5 +  a_3\tau_4 \tau_5) \nonumber \\
        +& 2  a_1\tau_1 (\tau_2 \tau_3 \tau_4 ( a_4\tau_2 \tau_3 + 
		a_3\tau_2 \tau_4 +  a_2 \tau_3 \tau_4)
		+ a_5(\tau_3^2 \tau_4^2 + \tau_2^2 (\tau_3^2 + \tau_4^2)) \tau_5 
		+ ( a_4(\tau_2^2 + \tau_3^2) \tau_4 \nonumber \\
        +& a_3\tau_3 (\tau_2^2 + \tau_4^2) + 
		a_2 \tau_2 (\tau_3^2 + \tau_4^2)) \tau_5^2) 
		+ \tau_2^2 (1 - 2 K \tau_3^2 - 2  a_4\tau_4^2 - 2 K \tau_4^2 + 
		a_4^2 \tau_3^2 \tau_4^2 \nonumber \\
		+& 2  a_4K \tau_3^2 \tau_4^2 + 
		K^2 \tau_3^2 \tau_4^2 
		+ 2  a_4 a_5\tau_3^2 \tau_4 \tau_5 
		+ (-2  a_5- 2 K +  a_5^2 \tau_3^2 + 
		2  a_5K \tau_3^2 \nonumber \\
        +& K^2 \tau_3^2 + ( a_4+  a_5+ K)^2 \tau_4^2) \tau_5^2 + 
		a_3^2 \tau_3^2 (\tau_4^2 + \tau_5^2)
		+ 2  a_3\tau_3 (\tau_3 (-1 + ( a_4+ K) \tau_4^2) \nonumber \\
		+& 
		a_5\tau_4^2 \tau_5 + (( a_5+ K) \tau_3 + 
		a_4\tau_4) \tau_5^2) + 
		a_2^2 (\tau_4^2 \tau_5^2 + \tau_3^2 (\tau_4^2 + \tau_5^2)) + 
		2  a_2 (-\tau_4^2 + (-1 + ( a_4+  a_5+ 
		K) \tau_4^2) \tau_5^2 \nonumber \\
		+& \tau_3^2 (-1 + ( a_3+  a_4+ 
		K) \tau_4^2 + ( a_3+  a_5+ 
		K) \tau_5^2))) + \tau_1^2 (1 - 2 K \tau_2^2 - 
		2  a_3\tau_3^2 \nonumber \\
        -& 2 K \tau_3^2 + a_3^2 \tau_2^2 \tau_3^2 + 
		2  a_3K \tau_2^2 \tau_3^2
		+ K^2 \tau_2^2 \tau_3^2
		+ 2  a_3 a_4\tau_2^2 \tau_3 \tau_4 \nonumber \\
        -& 2  a_4\tau_4^2 - 2 K \tau_4^2 +
		a_4^2 \tau_2^2 \tau_4^2 + 2  a_4K \tau_2^2 \tau_4^2 + 
		K^2 \tau_2^2 \tau_4^2 
		+ a_3^2 \tau_3^2 \tau_4^2 
		+2  a_3 a_4\tau_3^2 \tau_4^2 
		+  a_4^2 \tau_3^2 \tau_4^2 \nonumber \\
		+& 
		2  a_3K \tau_3^2 \tau_4^2 
		+ 2  a_4K \tau_3^2 \tau_4^2 + 
		K^2 \tau_3^2 \tau_4^2 + 
		2  a_5( a_4(\tau_2^2 + \tau_3^2) \tau_4 + 
		a_3\tau_3 (\tau_2^2 + \tau_4^2)) \tau_5 
		+ (( a_3\tau_3 + 
		a_4\tau_4)^2 \nonumber \\
		+&  a_5^2 (\tau_2^2 + \tau_3^2 + \tau_4^2)
		+
		K^2 (\tau_2^2 + \tau_3^2 + \tau_4^2) + 
		2 K (-1 +  a_3\tau_3^2 +  a_4\tau_4^2) + 
		2  a_5(-1 +  a_3\tau_3^2 +  a_4\tau_4^2 \nonumber \\
		+& 
		K (\tau_2^2 + \tau_3^2 + \tau_4^2))) \tau_5^2
		+
		a_2^2 \tau_2^2 (\tau_3^2 + \tau_4^2 + \tau_5^2)
		+
		a_1^2 (\tau_4^2 \tau_5^2 + \tau_3^2 (\tau_4^2 + \tau_5^2) + \tau_2^2 (\tau_3^2 + \tau_4^2 + \tau_5^2)) \nonumber \\
		+& 
		2  a_2 \tau_2 (\tau_3 \tau_4 ( a_4\tau_3 +  a_3\tau_4) \nonumber \\
		+& 
		a_5(\tau_3^2 + \tau_4^2) \tau_5
		+ ( a_3\tau_3 
		+
		a_4\tau_4) \tau_5^2 
		+ \tau_2 (-1 +  a_3\tau_3^2 + 
		K \tau_3^2 +  a_4\tau_4^2 + 
		K \tau_4^2 + ( a_5+ K) \tau_5^2)) + 
		2  a_1(-\tau_4^2 \nonumber \\
		+& (-1 + ( a_4+  a_5+ 
		K) \tau_4^2) \tau_5^2 
		+ \tau_2^2 (-1 +  a_2 \tau_3^2 + 
		a_3\tau_3^2 + K \tau_3^2 +  a_2 \tau_4^2 +  a_4\tau_4^2 + 
		K \tau_4^2 \nonumber \\
		+& ( a_2 +  a_5+ K) \tau_5^2)
		+ \tau_3^2 (-1 + ( a_3+
		a_4+ K) \tau_4^2 + ( a_3+  a_5+ K) \tau_5^2))) \nonumber \,.
	\end{empheq}

	\section{Subfiffusion with different $c$ and $n$}\label{sec_subdiff_c_n}
    
\begin{figure*}
			\centering
			\includegraphics[width=\textwidth]{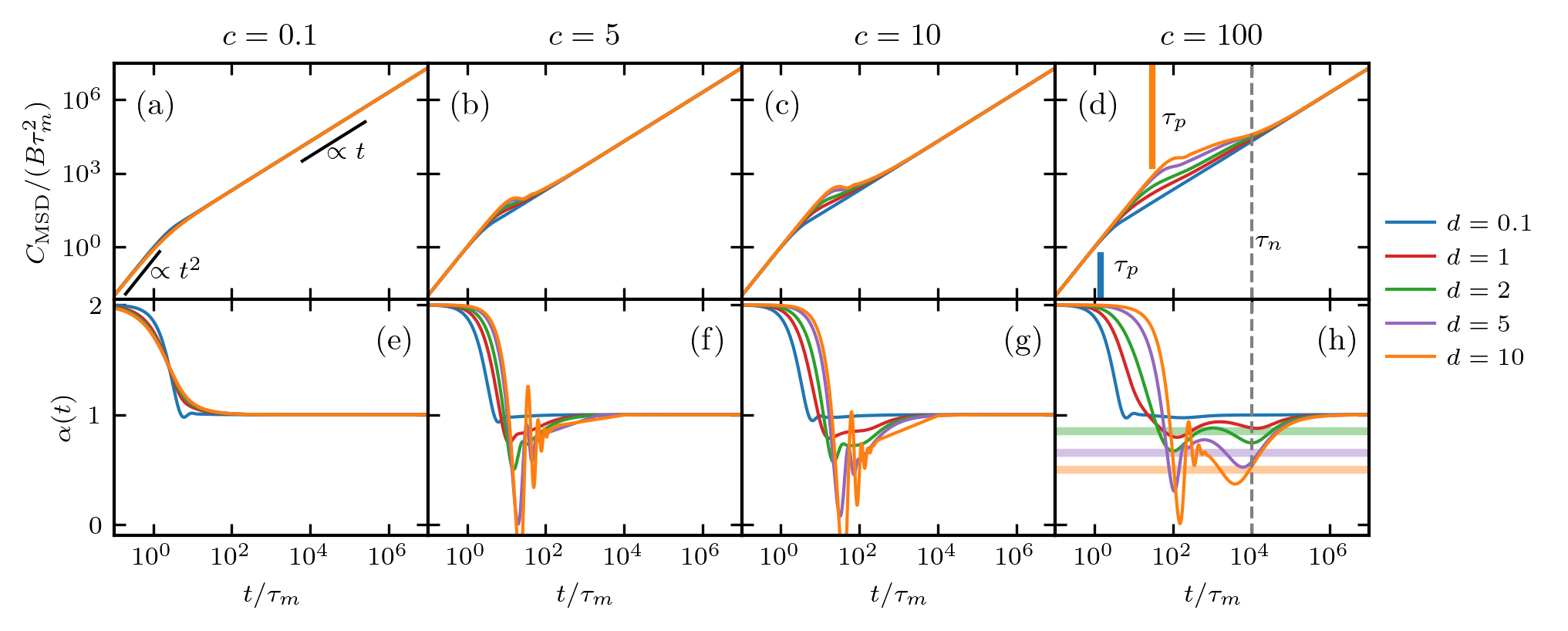}
			\caption{Analytical MSDs from Eq.~\eqref{eq_msd_general_expos} with $U(x)=0$ for different combinations of $c$ and $d$, defined in Eq.~\eqref{eq_cd} with $n=3$ exponential memory terms in the upper row and $\alpha(t)$ in the lower row.
			The first memory time scale is $\tau_1=\tau_m$ and the black lines in (a) indicate the limiting exponents $\alpha=1$ and $\alpha=2$.
			The vertical lines in (d) represent the persistence time $\tau_p$ given by Eq.~\eqref{eq_def_taup} in the main text.
			The thick horizontal lines in (g) and (h) represent the prediction of subdiffusive scaling $\alpha$ by Eq.~\eqref{eq_alpha_cd} in the respective color.}
			\label{fig_msd_free_alpha_n3}
		\end{figure*}

    In Fig.~\ref{fig_msd_free_alpha_n5} of the main text, we show the MSD and its time-dependent exponent $\alpha(t)$ for $n=5$ exponentials in the memory kernel Eq.~\eqref{eq_kernel_general_sum_expos}. The subdiffusive regime, resulting from certain combinations of friction amplitudes and time scales defined via Eq.~\eqref{eq_cd}, is prolonged by using more exponential memory components, since the longest memory time $\tau_n$ increases, as indicated by the end of the subdiffusive regime at $\tau_n$ in Fig.~\ref{fig_msd_free_alpha_n3}d,h and in the main text Fig.~\ref{fig_msd_free_alpha_n5}c,f,i.
    The running integral $G(t)$ can be used to predict the persistence time $\tau_p$, as discussed in the main text. We determine $\tau_p$ by iterating $\tau^p_{j} = 1/ G(\tau^p_{j-1})$ until reaching a fixed point; this can be done independently of whether $G(t)$ is extracted from data or whether one uses an analytical expression.
    The MSD curves in Fig.~\ref{fig_msd_free_alpha_n3} are obtained by our analytical result, derived in Sec.~\ref{sec_msd_expo_derivation}. However, for the MSDs from simulations in a double well potential, shown in the main text Figs.~\ref{fig_dw_msd_nomem} and \ref{fig_dw_msd_mem}, results are obtained by averaging over 1000 trajectories, each of length $10^8$ time steps, where we first perform a time average to obtain the MSD for a single trajectory and then average over the trajectory ensemble.
    The time steps in the simulations are given for Fig.~\ref{fig_dw_msd_nomem}a and b as $\Delta=0.1\tau_m$ and $\Delta=\tau_m$, respectively. Here, the simulations are performed by solving Eq.~\eqref{eq_le_wom} with a fourth-order Runge-Kutta integrator, and initial conditions for each simulation are drawn from the respective Boltzmann distribution.
    In Fig.~\ref{fig_dw_msd_mem}, the simulation time step is $\Delta=100\tau_m$, and we use the Runge-Kutta integrator to solve the Markovian embedding that describes the GLE Eq.~\eqref{eq_gle_wom} with the kernel Eq.~\eqref{eq_kern_expo} \cite{noauthor_advances_2012,brunig_time-dependent_2022}.

\begin{figure*}
		\centering
		\includegraphics[width=\textwidth]{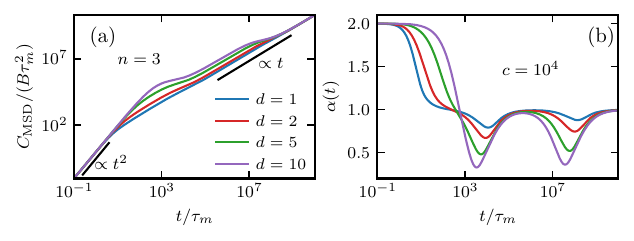}
		\caption{(a) Analytical MSDs according to Eq.~\eqref{eq_msd_general_expos} without external potential, $U(x)=0$ for $n=3$, $c=10^4$ and $\tau_1=\tau_m$ and (b) the respective time-dependent exponent $\alpha(t)$.
		Black lines in the MSD indicate the asymptotic scaling behavior.
		}
		\label{fig_msd_alpha_bigc_n3}
	\end{figure*}

\begin{figure*}
		\centering
		\includegraphics[width=\textwidth]{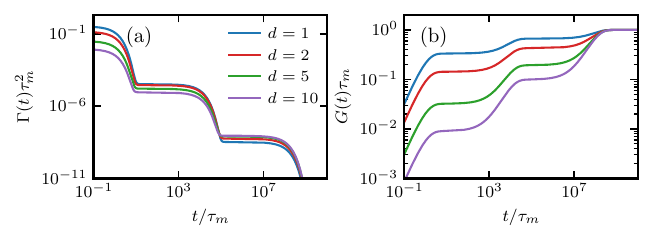}
		\caption{(a) Multi-exponential kernel $\Gamma(t)$ for $n=3$, $c=10^4$ and $\tau_1=\tau_m$ and (b) the respective running integral $G(t)$.}
		\label{fig_kern_int_bigc_n3}
	\end{figure*}

    Minima of $\alpha(t)$ can occur separately if the ratio of the temporal scales $c$ becomes large, as demonstrated in Fig.~\ref{fig_msd_alpha_bigc_n3} for $n=3$.
    One can see an intermediate diffusive regime between the subdiffusive regimes in Fig.~\ref{fig_msd_alpha_bigc_n3}a, which exhibits a larger diffusivity for higher values of $d$, i.e., with less momentary friction $G(t)$ on this time scale.
    The diffusivity of this intermediate regime is determined by the intermediate plateau of the integrated friction $G(t)$, seen in Fig.~\ref{fig_kern_int_bigc_n3}b.
    The corresponding time-dependent exponent $\alpha(t)$ in Fig.~\ref{fig_msd_alpha_bigc_n3}b exhibits two distinct minima separated by an intermediate diffusive regime with $\alpha=1$.
    For large separation of the memory times, the shortest memory time acts similarly to a delta-contribution at time zero.
    The friction amplitude of this shortest memory contribution compared to the total friction, $\gamma_1/G(\infty)$, decreases as $d$ increases, as seen by decreasing intermediate plateau values in $G(t)$ in Fig.~\ref{fig_kern_int_bigc_n3}b, which leads to increasingly long ballistic regimes for increasing $d$.
    For $d=1$, the end of the ballistic regime and the first minimum in $\alpha(t)$ are well separated so that there is another diffusive regime in between, meaning a total of three diffusive regimes with $\alpha=1$, as seen for the blue line in Fig.~\ref{fig_msd_alpha_bigc_n3}b.

	\makeatletter\@input{aux_file.tex}\makeatother